\documentclass[fleqn,usenatbib]{mnras}

\usepackage[T1]{fontenc}
\usepackage{ulem}
\DeclareRobustCommand{\VAN}[3]{#2}
\let\VANthebibliography\thebibliography
\def\thebibliography{\DeclareRobustCommand{\VAN}[3]{##3}\VANthebibliography}

\usepackage{graphicx}	
\usepackage{amsmath}	
\usepackage{array}
\usepackage{mathabx}
\usepackage{soul}
\usepackage{xcolor}
\usepackage{adjustbox}
\usepackage{placeins}
\usepackage{orcidlink}

\newcommand{\ond}{Ond\v{r}ejov}

\title[A Detailed Study of DD CrB]{A Detailed Study of the Physical and Orbital Characteristics, and Eclipse Timing Variations of the Post Common Envelope Binary DD CrB}
\author[Ba\c{s}t\"urk et al.]{%
Ba\c{s}t\"urk, \"O.$^{1,2}$\thanks{E-mail: obasturk@ankara.edu.tr}\orcidlink{0000-0002-4746-0181},
Sertkan, E.$^{3}$\orcidlink{0009-0001-9812-2780},
Akar, F.$^{3}$\orcidlink{0000-0003-4419-2908},
Munari, U.$^{4}$\orcidlink{0000-0001-6805-9664},
G\"uler, B.$^{3}$\orcidlink{0009-0005-5049-5720},
Esmer, E.M.$^{1,2,5}$\orcidlink{0000-0002-6191-459X},
\newauthor
Selam, S.O.$^{1,2}$\orcidlink{0000-0002-4953-4818},
Kutluay, A.C.$^{6,3}$\orcidlink{0009-0007-0502-7359},
Wolf, M.$^{7}$\orcidlink{0000-0002-4387-6358},
Zasche, P.$^{7}$\orcidlink{0000-0001-9383-7704},
Kučáková, H.$^{7}$\orcidlink{0000-0002-1330-1318}
Zejda, M.$^{8}$\orcidlink{0000-0001-6231-3350},
\newauthor
\c{S}im\c{s}ir, \"O$^{3}$\orcidlink{0009-0007-8172-6602}\\
\\
$^{1}$\,Ankara University, Faculty of Science, Astronomy \& Space Sciences Department, Tando\u{g}an, TR-06100, Ankara, T\"urkiye\\
$^{2}$\,Ankara University, Astronomy and Space Sciences Research and Application Center (Kreiken Observatory),\\
{\.I}ncek Blvd., TR-06837, Ahlatlıbel, Ankara, T\"urkiye\\
$^{3}$\,Ankara University, Graduate School of Natural \& Applied Sciences, Astronomy \& Space Sciences Department,\\ Ziraat Mahallesi {\.I}rfan Ba\c{s}tu\u{g} Caddesi, D{\i}\c{s}kap{\i}, TR-06110 Alt{\i}nda\u{g} / Ankara, T\"urkiye\\
$^{4}$\,INAF National Institute of Astrophysics, Astronomical Observatory of Padova, 36012 Asiago (VI), Italy\\
$^{5}$\,Department of Physics and McDonnell Center for the Space Sciences, Washington University, St. Louis, MO 63130, USA\\
$^{6}$\,Astrophysics Group, Keele University, Staffordshire, ST5 5BG, UK\\
$^{7}$\,Astronomical Institute, Faculty of Mathematics and Physics, Charles University, Praha 8, Czech Republic\\ 
$^{8}$\,Department of Theoretical Physics and Astrophysics, Masaryk University, Brno, Czech Republic\\
}
\date{Accepted XXX. Received YYY; in original form ZZZ}

\pubyear{2015}

\begin{document}
\label{firstpage}
\pagerange{\pageref{firstpage}--\pageref{lastpage}}
\maketitle

\begin{abstract}

We present an in-depth analysis of the eclipsing binary DD~CrB, composed of a B-type subdwarf primary and an M-type main-sequence secondary, with the main goal of investigating its eclipse timing variations (ETVs). Our new multi-color photometric observations, radial velocity measurements, and precise eclipse timings from TESS allow us to constrain the system parameters. The R{\o}mer delay between primary and secondary minima yields a mass ratio of $q = 0.299 \pm 0.009$, enabling robust simultaneous modeling of the light and radial velocity curves with {\sc phoebe} 2.17. By fixing the albedo of the secondary to its maximum physically plausible value (A$_2 = 1.0$), despite the degeneracy between albedo, surface temperature, and radius, we obtained a satisfactory fit, resulting in a significantly lower temperature ($T_2 \sim 2360$ K) and a radius ($R_2 \sim 0.16$ R$_\odot$) in agreement with literature values. Using the total mass of the components and the orbital size derived from this modeling, we interpret the ETVs and find them best explained by a Jupiter-mass tertiary companion on a $\sim13$-year orbit in all competing models, while the eccentric (e $\sim0.46$) models perform better in terms of fit statistics.
\end{abstract}

\begin{keywords}
methods: data analysis – techniques: photometric – stars: fundamental parameters – binary stars: individual: DD~CrB – planetary systems.
\end{keywords}



\section{Introduction}
\label{sec:introduction}

Eclipsing post-common envelope binaries (PCEB) belong to a small group of close binary stars that are very important for resolving the very short-lived common envelope phase of stellar evolution and its consequences. Many such systems consist of very hot subdwarf stars and cool M-type stars with a typical difference between primary and secondary surface temperatures of about 30,000~K. Their characteristic light curves with a deep and narrow primary eclipse and a strong reflection effect are a clear sign allowing for the simple identification  of these unique objects. Short orbital periods of about 2-3 hours are very sensitive to any changes possibly caused by magnetic activity and/or the presence of an unseen, potentially planetary-mass third body, which may have survived the common envelope ejection (first generation) or formed from the ejecta (second generation) \citep{Zorotovic2013,bear_soker2014}. Such a gravitationally bound third body can even survive later evolutionary stages of both companions \citep{columba2023, ledda2023} although its orbital and physical properties are subject to change (hybrid scenario).   

DD~CrB (also NSVS 7826147, 2MASS J15334944+3759282)
is an eclipsing PCEB a B-type subdwarf primary (sdB) and an M-dwarf secondary (dM) in a short-period orbit (P$_{\rm orb} = 0.16177$~d). The large temperature contrast and close separation between the components give rise to a strong reflection effect from the surface of the cool secondary, modulated by the orbital phase. Such systems are also referred to as HW~Vir-type binaries, after the prototype of this class. The primary (deeper) eclipse occurs when the hotter sdB component is occulted at the phase of minimum reflection by the cooler dM star, whose irradiated hemisphere is directed away from the observer. The secondary minimum, on the other hand, occurs half an orbital cycle later at the phase of maximum reflection when the illuminated side of the secondary is hidden behind the much hotter primary due to the high orbital inclination. Because the irradiated hemisphere of the secondary is externally heated and thus emits differently from an unirradiated M~dwarf, the reflection effect is also strongly wavelength-dependent. Consequently, determination of the system’s physical parameters depends critically on accurate modelling of this reflection.

DD~CrB was identified as an eclipsing binary containing a hot subdwarf~B (sdB) primary by \citet{kelly_shaw_2007}, and was subsequently studied in detail for the first time by \citet{for2010}. Using low-resolution spectroscopic observations obtained with the 2.3\,m Bok Telescope at Kitt Peak, they derived the effective temperature of the primary component as 29,230 K. They further determined the system’s fundamental physical and orbital parameters by modelling its multi-band ($B$, $V$, $R$, $I$) light curves and radial velocity curve with the Wilson–Devinney code \citep{wilson_devinney1971}. A satisfactory fit to the reflection effect could only be achieved when the bolometric albedo of the secondary was allowed to exceed unity, an unphysical value that likely compensates for model limitations. Because DD~CrB is a single-lined spectroscopic binary, their solutions with different assumed mass ratios and orbital inclinations yielded a range of primary masses. Among these, the model with q$ = 0.301$ provided the most physically consistent results, giving a primary mass of M$_{1} = 0.376 \pm 0.055~M_{\odot}$ within the typical range for sdB stars, though smaller than the canonical value of $\sim 0.47~M_{\odot}$ \citep{rojas2024}. The corresponding orbital inclination was found to be i$ = 86\fdg6 \pm 0\fdg2$.
 
\citet{zhu_qian_2010} performed a light-curve analysis of DD~CrB using a single $R$-band dataset. In their modelling, they adopted the primary temperature derived spectroscopically by \citet{for2010} and fixed the secondary’s temperature to 3000~K, adopting the mass ratio as q$ = 0.30$ from \citet{for2010}. As in the earlier study, their solution required adjusting the secondary’s bolometric albedo to values exceeding unity; they obtained A$_{2} = 2.1 \pm 0.2$.

Later, \citet{lee2017} modelled the $B$- and $V$-band light curves of DD~CrB simultaneously with the radial velocity curve from \citet{for2010}. They performed a q-search to constrain the system’s mass ratio and identified a global minimum at q$ = 0.28$, which they adopted as an initial value but allowed it to vary during the modelling. The effective temperature of the primary was fixed to the value derived by \citet{for2010}. Their best-fitting solution yielded q$ = 0.2802 \pm 0.0049$, significantly smaller than the value obtained by \citet{for2010}, leading to a primary mass of M$_{1} = 0.442 \pm 0.012~M_{\odot}$, much closer to the canonical value for sdB stars. They achieved a consistent model for both bands by fixing the bolometric albedos of both components to 1.0 and derived a surface temperature of approximately $3100$~K for the secondary.

\citet{backhaus2012} published the first O–C diagram for DD~CrB, which showed no evidence of variability based on the accumulated eclipse timing measurements available at the time. \citet{lohr2014} also found a constant orbital period and updated the system’s linear ephemeris. \citet{zhu2015a} reported a periodic variation in the O–C diagram based on a longer baseline of eclipse timings, which they later \citep{zhu2015b} attributed to a light-time effect (LiTE) caused by a companion with a minimum mass of 4.7~M$_{\mathrm{Jup}}$ on a circular orbit with a radius of 0.64~au. \citet{lee2017} however, they found no evidence for such periodicity in their O–C diagram covering a 12-year timespan. \citet{pulley2018} noted that the scatter in their O–C data was larger than the $\sim$3.5 s amplitude expected from the LiTE signal with the parameters proposed by \citet{zhu2015b}. \citet{wolf2021} analyzed 284 reliable eclipse timings, including TESS data from Sector 24, and found the orbital period to vary with a periodicity of $10.5 \pm 0.4$ yr. They attributed this variation to a LiTE caused by an additional gravitationally bound object with a minimum mass of 1.36~M$_{\mathrm{Jup}}$, adopting the physical parameters of the binary components from \citet{lee2017}. Finally, \citet{pulley2025} added 51 new minima timings to their work in 2018 \citep{pulley2018} and updated their analysis and proposed that the eclipse timings after 2017 do not follow the LiTE model suggested by \citet{wolf2021}.

With the availability of four additional TESS sectors and new high-precision ground-based photometric and spectroscopic observations, we revisit the DD~CrB system to update its fundamental parameters with a more tightly constrained mass ratio (q). We also construct an updated O–C diagram spanning a longer observational baseline to investigate the presence of any eclipse timing variations (ETVs). In Section~\ref{sec:observations}, we describe our photometric (Section~\ref{subsec:photometric_observations}) and spectroscopic (Section~\ref{subsec:rv_obs}) observations in detail. These data are analyzed using the second version of the {\sc phoebe} software package \citep{conroy2020,conroy2021}, as outlined in Section~\ref{subsec:lc_rv_analysis}, to derive the system’s fundamental physical and orbital parameters. The construction and analysis of the O–C diagram are presented in Section~\ref{sec:etv_analysis}, while the results and their implications are discussed in Section~\ref{sec:discussion}.
 
\section{Observations and Data Reduction}
\label{sec:observations}

\subsection{Photometric Observations}
\label{subsec:photometric_observations}

DD~CrB was observed photometrically using both ground-based and space-based facilities. Ground-based observations were carried out over multiple nights between May 2021 and May 2025 using five different telescopes located at four observatory sites. These include the 1.0\,m Telescope (TUG100) at the Bak{\i}rl{\i}tepe site of the Turkish National Observatories, the 80\,cm Prof. Dr. Berahitdin Albayrak Telescope (T80) and the 35~cm T35 telescope at the Ankara University Kreiken Observatory (AUKR), the 62\,cm telescope at the Krav\'{i} Hora Observatory (Brno, Czech Republic), the 80\,cm Masaryk University telescope at the \v{Z}d\'{a}nice Observatory (\v{Z}d\'{a}nice, Czech Republic), and the 65~cm Mayer telescope at \ond\ Observatory (\ond, Czech Republic). The observations were conducted through various filters to provide photometric coverage across a broad spectral range, and exposure times were optimized to achieve a high signal-to-noise ratio. A log of our photometric observations, including details on the telescope, filter, exposure time, and photometric scatter, is provided in Table~\ref{tab:logphotometry}. In addition, we made use of space-based photometric data obtained by the Transiting Exoplanet Survey Satellite (TESS).

The ground-based images were corrected for dark current, bias, and flat-field effects, and differential photometry was subsequently performed using the \textsc{AstroImageJ} software package \citep{aij2017}. Light curves were extracted using an optimal set of comparison stars, dynamically selected to minimize scatter based on their proximity to the target, brightness, and photometric stability. When variations in airmass introduced noticeable trends, an airmass detrending correction was applied during the reduction process.

\subsubsection{TUG100 Observations}
\label{subsubsec:TUG100}
Photometric observations of DD~CrB were carried out using the TUG100 telescope at the Turkish National Observatories in Bak{\i}rl{\i}tepe, Antalya (altitude 2500\,m). The telescope has a 1.0\,m primary mirror with an f/10 focal ratio and is equipped with a cryo-cooled SI~1100 back-illuminated CCD camera featuring a $4096 \times 4096$ pixel array with 15\,$\mu$m pixels, yielding an effective field of view of $21 \times 21$~arcmin in unbinned mode. The detector readout time is approximately 45~s; therefore, a $2\times2$ binning mode was applied to reduce it to roughly 15~s, allowing for a higher cadence in the resulting light curves. Observations were conducted using the SDSS filter set.

\subsubsection{T80 Observations}
\label{subsubsec:t80}
The target object was observed using the 80\,cm Prof. Dr. Berahitdin Albayrak Telescope (T80) at the AUKR, located at an altitude of approximately 1250\,m. The telescope has an 80\,cm diameter primary mirror with an f/7 focal ratio and is equipped with an Apogee Alta U47+ back-illuminated CCD camera with a 1024$\times$1024 pixel array and a pixel size of 13\,$\mu$m. A focal reducer provides a field of view of 11.84$\times$11.84 arcminutes. Observations were conducted using the SDSS filter set.

\subsubsection{T35 Observations}
\label{subsubsec:t35}
A single additional observation was conducted using the 35\,cm Yrd. Do\c{c}. Dr. Zekeriya M\"uyessero\u{g}lu Telescope (T35) at the AUKR. The telescope has a 35\,cm diameter primary mirror with an f/10 focal ratio and is equipped with a QSI 660 front-illuminated CCD camera with a resolution of $2758\times2208$ pixels and a pixel size of 4.54\,$\mu$m, providing a field of view of $12.10\times9.96$ arcminutes. A $2\times2$ binning mode was applied to decrease the exposure and readout time, considering the telescope aperture. The observation was carried out using a Bessel-$R$ filter.

\subsubsection{Photometric Observation in Brno Observatory}
\label{subsubsec:masaryk_obs}
Photometric data of DD~CrB were acquired using the 62\,cm telescope at the observatory located at Kraví Hora in Brno (Czech Republic). The telescope with a focal length is 2750 mm was equipped with CCD camera Moravian Instruments G4-16000 with a resolution 4096 $\times$ 4096 pixels and a pixel size 9\,$\mu$m. A $2\times2$ binning mode was applied. During the observations, Johnson \textit{VR} filters were used.

\subsubsection{Photometric Observation in \v{Z}d\'{a}nice Observatory}
\label{subsubsec:zdanice_obs}
Photometric observations were obtained using the ASA telescope AZ800 of Masaryk University, located at the Old\v{r}ich Kot\'{i}k Observatory and Planetarium in \v{Z}d\'{a}nice, Czech Republic. The mirror of the telescope has a diameter of 800 mm and a focal ratio of f/6.8. During DD~CrB observations, two cameras from Moravian Instruments were used. The first one, CCD camera G4-16000 with a resolution $4096\times4096$ pixels and a pixel size 9\,$\mu$m was used in $2\times2$ binning mode with \textit{VR} filters. The second camera, CMOS C5A-150M with a resolution of $14208\times10656$ pixels and $3.76\times3.76$\,$\mu$m was used in binning mode $4\times4$ with Sloan \textit{gr} filters.  

\bigskip
\subsubsection{Photometric Observation in \ond\ Observatory}
\label{subsubsec:ondrejov_obs}
Since  2021 photometric observations of DD~CrB continued at \ond\ observatory, Czech Republic, as a part of long-term photometric program of monitoring period changes of selected PCEBs. 
The Mayer 0.65-m ($f/3.6$) reflecting telescope, G2-3200 CCD camera, and photometric {\it VR} filters were used. An exposure time of 30~seconds and binning $2\times2$ was usually applied.

\begin{table}
  \centering
   \caption{Log of photometric observations of DD~CrB. Observations marked with an asterisk (\textsuperscript{*}) were used in the light curve analysis, as they provide full orbital phase coverage and lower photometric scatter.}  
  \begin{tabular}{ccccc}
    \hline
    Date of Obs. & Telescope & Filter & Exp. Time & $\sigma_{\rm phot}$ \\
     &           &         & (sec)     & (mmag) \\
    \hline
    2023 04 18\textsuperscript{*} & TUG100 & SDSS-i'   & 40  & 3.408 \\
    2023 05 15\textsuperscript{*} & TUG100 & SDSS-z'   & 110 & 6.039 \\
    2023 02 21\textsuperscript{*} & TUG100 & SDSS-r'   & 80  & 2.067 \\
    2024 04 04\textsuperscript{*} & TUG100 & SDSS-u'   & 60  & 6.183 \\
    \hline
    2022 07 04\textsuperscript{*}  & T80  & SDSS-g'   & 45  & 1.985 \\
    2022 07 06                     & T80  & SDSS-r'   & 60  & 2.370 \\
    2023 07 02                     & T80  & SDSS-i'   & 70  & 6.699 \\
    2024 02 28                     & T80  & SDSS-i'   & 40  & 10.877 \\
    2024 04 09                     & T80  & SDSS-g'   & 45  & 3.003 \\
    2024 04 14                     & T80  & SDSS-z'   & 150 & 5.674 \\
    2024 06 04                     & T80  & Clear     & 35  & 2.430 \\
    2024 08 20                     & T80  & SDSS-r'   & 40  & 5.407 \\
    \hline
    2024 08 20                     & T35  & Bessel-R  & 120 & 16.457 \\
    \hline
    2021 05 15                     & Brno & Johnson-R  & 70  & 29.054 \\
    2022 02 13                     & Brno & Johnson-R  & 70  & 12.385 \\
    \hline
    2023 04 22                     & Ždánice & Johnson-R & 70  & 4.810 \\
    2023 05 04                     & Ždánice & Johnson-V & 80 & 11.774 \\
    2024 05 29                     & Ždánice & Johnson-R & 70  & 5.343 \\
    2025 05 03                     & Ždánice & Sloan-g  & 200 &  9.496 \\
    2025 05 03                     & Ždánice & Sloan-r  & 200 & 17.571 \\
    \hline
    2020 09 15                     & \ond    & Johnson-R & 30 &   7.0 \\
    2021 03 25                     & \ond    & Johnson-R & 30 &  10.0  \\
    2021 04 20                     & \ond    & Johnson-R & 30 &   5.0  \\
    2021 06 01                     & \ond    & Johnson-R & 30 &   5.0  \\
    2021 10 18                     & \ond    & Johnson-R & 60 &  12.0  \\ 
    2022 04 19                     & \ond    & Johnson-R & 30 &   8.5  \\ 
    2022 06 16                     & \ond    & Johnson-R & 30 &   8.0  \\ 
    2023 03 18                     & \ond    & Johnson-R & 30 &   7.0  \\ 
    2023 03 28                     & \ond    & Johnson-R & 30 &   8.0  \\ 
    2024 04 18                     & \ond    & Johnson-R & 60 &   4.5  \\  
    2024 04 18                     & \ond    & Johnson-V & 60 &   5.0  \\ 
    2025 03 19                     & \ond    & Johnson-R & 30 &   4.5  \\
    2025 06 11                     & \ond    & Johnson-R & 30 &   4.0  \\  
    \hline
  \end{tabular}
  \label{tab:logphotometry}
\end{table}

\subsubsection{TESS Observations}
\label{subsubsec:tess}
TESS observed DD~CrB (TIC~148785530, TESS magnitude 13.205) during sectors 24–51 in short-cadence mode, using a red-optical bandpass centered near the Cousins~$I_c$ band. We downloaded the photometric products from the Barbara A.~Mikulski Archive for Space Telescopes (MAST\footnote{\url{https://archive.stsci.edu}}) at the Space Telescope Science Institute (STScI) of NASA. Short-cadence light curves processed by the Science Processing Operations Center (SPOC) pipeline \citep{jenkins2016} were retrieved using the SAP (Simple Aperture Photometry) data products via the \texttt{astroquery} package \citep{astroquery2019}.

Short-cadence observations were obtained in Sectors 24, 50, and 51. In Sector 24, the data were acquired with Camera 1 and CCD 2 (sector midpoint April 2020); in Sector 50, with Camera 3 and CCD 2 (April 2022); and in Sector 51, with Camera 3 and CCD 1 (May 2022). In addition, DD~CrB was observed in Sectors 77 and 78, for which no short-cadence data are available. To extend the temporal baseline of the photometric dataset and to provide additional eclipse timing measurements at later epochs, we therefore made use of the Full Frame Image (FFI) light curves produced by the Quick-Look Pipeline (QLP; \citealt{huang2020}), which provides aperture photometry extracted directly from the TESS FFIs with an effective cadence of 200 s. In Sector 77, DD~CrB was observed with Camera~1 and CCD~1 (April~2024), while in Sector 78 the observations were obtained with Camera~1 and CCD~2 (May~2024).

\begin{figure}
    \centering
    \includegraphics[width=1\columnwidth]{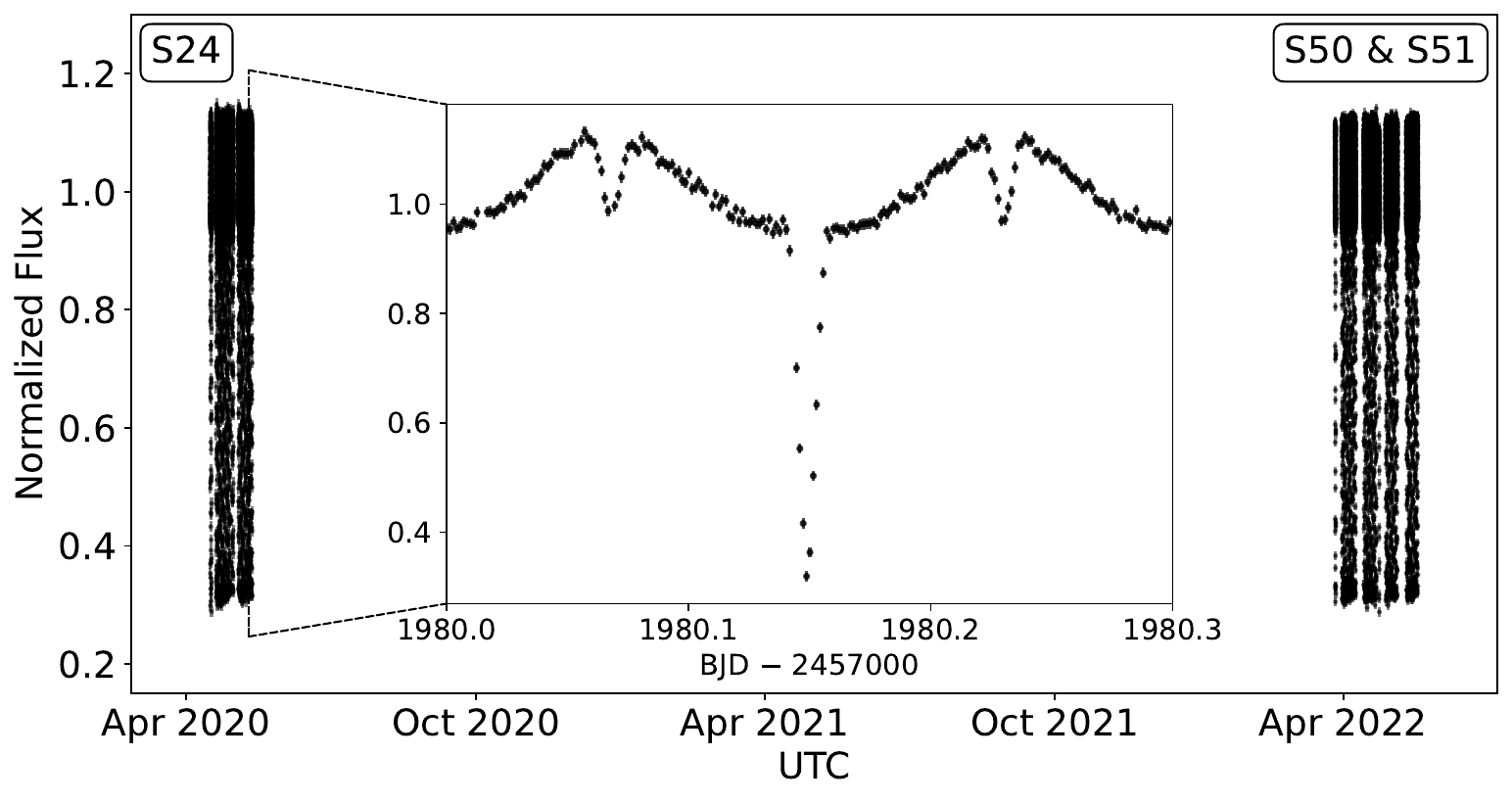}
    \caption{TESS short-cadence observations of DD~CrB spanning 2020--2022. The zoomed-in region highlights a segment of the time-series light curve from Sector~24.}
    \label{fig:ddcrb_tess_summary}
\end{figure}

The data were filtered according to the quality flags. Due to the non-continuous nature of the observations, each sector was divided into segments to which second-degree polynomial fits were applied, followed by normalization to the median flux level. An example of the processed short-cadence TESS light curves is shown in Figure \ref{fig:ddcrb_tess_summary}.

After reducing the photometric data, mid-eclipse times for all observed minima were determined using the Kwee-van Woerden (KvW) method, as implemented in the \textsc{Xtrema} software \citep{xtrema2015}. The secondary eclipse minima were found to exhibit larger scatter, likely reflecting the intrinsic limitations of the KvW method in accurately determining their timings. To mitigate these uncertainties and enhance the robustness of the eclipse timing analysis, only the primary eclipse minima were considered in this study.

\subsubsection{Archival Eclipse Timing Data}
\label{subsubsec:minima_data}
In addition to the newly obtained ground-based and space-based photometric observations, eclipse timing data were compiled from the literature \citep{zhu_qian_2010,for2010,lee2017,wolf2021,backhaus2012}, as well as from publicly available archival light curves provided by the American Association of Variable Star Observers (AAVSO\footnote{\url{https://www.aavso.org}}) and the VarAstro database\footnote{\url{https://var.astro.cz}}. In several cases, published eclipse timings were available for the same epochs; however, when the original light curves were accessible, the eclipse minima were re-determined directly from the photometric measurements using a uniform procedure. This approach ensured internal consistency across the dataset and enabled the inclusion of additional minima for which no eclipse timings had been previously reported.

The dataset used for the ETV analysis includes 128 primary eclipse timings from the literature, 17 derived from AAVSO light curves, 448 measured from five TESS sectors and 39 newly determined in this study. All eclipse times were converted to dynamical barycentric Julian days (BJD$_{\mathrm{TDB}}$) prior to the analysis. We attributed the median of all error bars to the minima timings without associated uncertainties.

\subsection{Radial Velocity Observations}
\label{subsec:rv_obs}
\subsubsection{Spectroscopic Observations in Asiago Observatory}
\label{subsubsec:asiago_obs}
Spectroscopic observations of DD~CrB have been obtained with the 1.22\,m (+ B\&C spectrograph) and 1.82-m (+ REOSC \'Echelle spectrograph) telescopes of the Asiago Observatory. Table~\ref{tab:logspectra} provides a log of the spectroscopic observations of DD~CrB and the derived heliocentric radial velocities, which we converted to dynamical barycentric time by correcting for the light time effect between the heliocenter and barycenter using the relevant functions of the {\sc astropy} package \citep{astropy2013,astropy2018}.  The longest exposure times do not exceed 6.4\% of the orbital period, short enough to avoid introducing any noticeable phase-smearing.

\begin{figure}
\includegraphics[width=0.99\columnwidth]{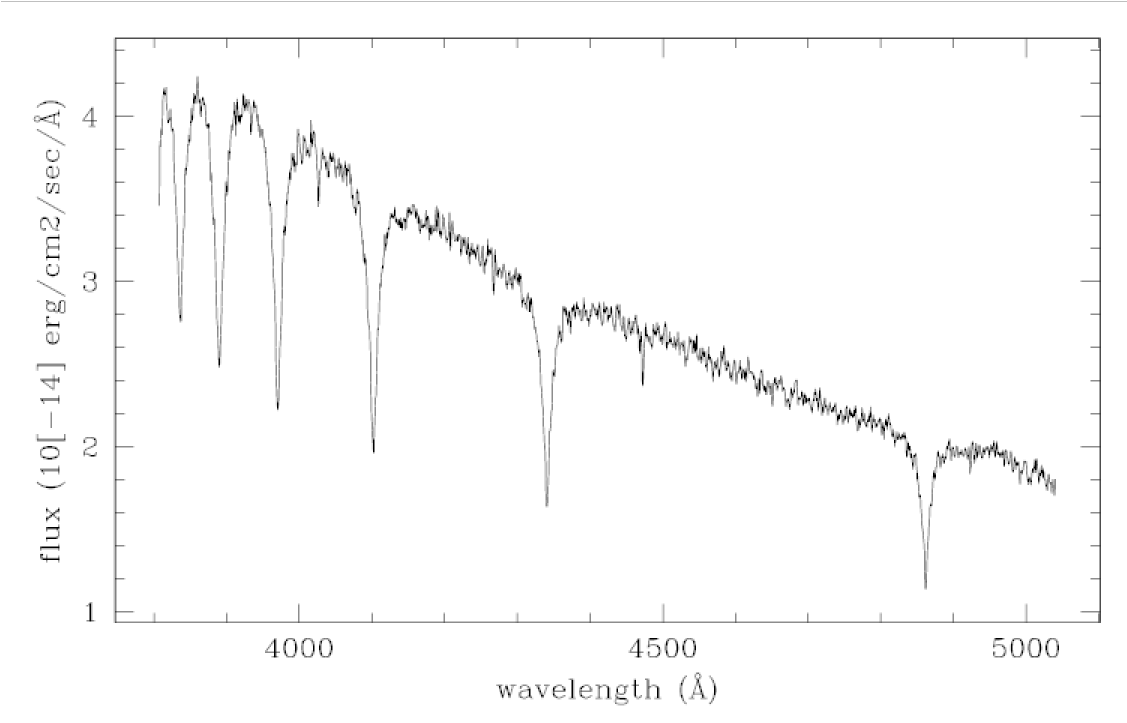}
    \caption{A sample spectrum covering 3810-5040 {\AA} region acquired with the 1.22\,m Asiago Telescope and B\&C spectrograph attached on its focal plane.}
    \label{fig:dd_crb_spec}
\end{figure}

The B\&C spectrograph mounted on the Asiago 1.22\,m telescope uses
an ANDOR iDus DU440A camera with an E2V 42-10 back illuminated CCD
($2048\times512$ array, 13.5 $\mu$m pixel size) of enhanced UV sensitivity.  For the observations of DD~CrB we selected a 1200 ln/mm grating blazed to 5000~\AA, which provides a dispersion of 0.59~\AA/pix and a recorded interval of 1210~\AA, covering from hydrogen H9 (3835~\AA) to HeI 5016~\AA. For wavelength calibration, we selected a FeArNe hollow-cathode lamp.  We provide a sample spectrum acquired with this setup in Figure \ref{fig:dd_crb_spec}.

The REOSC \'Echelle spectrograph of the Asiago 1.82m telescope feeds light to an Andor DW436-BV camera, which houses an E2V CCD42-40 AIMO back illuminated CCD as a detector ($2048\times2048$ array, 13.5 $\mu$m pixel size).  We adopted a 2-arcsec slit width, providing a resolving power of 20,000.  The 3600--7300~\AA\ interval is covered in 32 orders, without inter-order gaps. A Thorium hollow-cathode lamp is used for wavelength calibration, combining exposures taken immediately before and after the science spectrum.
\FloatBarrier
Observations at both telescopes involve the acquisition of dark and flat frames, while bias is derived from the over-scan regions framing all images. All reduction steps are performed in IRAF following the standard procedures outlined in \citet{2000iasd.book.....Z}, including sky background subtraction and wavelength calibration. The long slits at both spectrographs allow for a careful evaluation of the sky back-ground at the position of the science object.

The radial velocity of DD~CrB has been derived via cross-correlation with the {\sc RV/fxcor} package in IRAF.  For the \'echelle observations, we adopted as template a synthetic spectrum with $T_{\rm eff}$=24,000~K, $\log g$=5.0, [Fe/H]=0.0, $\zeta$=2.0 km/s, $v \sin i$=10~km/s taken from the \citet{2005A&A...442.1127M} synthetic library, which has been computed at the same 20,000 resolving power of the observed spectra. As template the cross-correlation of B\&C observations we used the spectra of HR 8634 (spectrum B8.5IV, Gaia DR3 RV=9.23 km/s) and HR 8967 (spectrum B9V, Gaia DR3 RV=5.93 km/s) obtained the same night and with the same spectrograph set-up as for DD~CrB \citep{gaia2016,gaia2021a,gaia2021b}. We attributed 5 km/s as an error bar to our RV measurements with the 1.88\,m telescope and 2.5 km/s to those with the 1.22\,m telescope of the Asiago Observatory, based on previous measurements with the instruments for stars of this magnitude and spectral type.

\begin{table*}
\caption{Log of the spectroscopic observations with the 1.22 and 1.82\,m telescopes in Asiago observatory.}
\begin{tabular}{ccrccrccc}
\hline\hline
&&\\
Date       &  UT   & (BJD$_{\rm TDB}$)${_{\rm mid.}}$ &  log N       &  t${_{\rm exp}}$  &  RV & Template &Asiago\\
           &       &        &                   & (sec.) & (km/s) & & telescope\\
&& \\
\hline
&& \\
2024 09 19 & 19:26:15  &   2460573.30810 &   72348   & 900 & -78.8 & synthe  & 1.82m+Echelle \\  
2024 09 19 & 19:36:42  &   2460573.31536 &   72349   & 140 & -73.4 & synthe  & 1.82m+Echelle \\  
2024 09 19 & 20:10:27  &   2460573.33879 &   72353   & 600 & -33.8 & synthe  & 1.82m+Echelle \\  
2024 09 21 & 19:37:05  &   2460575.31555 &   72367   & 900 &  66.3 & synthe  & 1.82m+Echelle \\  
2024 10 18 & 19:02:34  &   2460602.29101 &   72461   & 600 & -24.5 & synthe  & 1.82m+Echelle \\  
2024 10 18 & 19:14:04  &   2460602.29900 &   72462   & 600 &   4.9 & synthe  & 1.82m+Echelle \\  
2024 10 18 & 19:25:33  &   2460602.30696 &   72463   & 600 &   7.6 & synthe  & 1.82m+Echelle \\  
2024 11 11 & 17:13:38  &   2460626.21543 &   72498   & 900 & -63.7 & synthe  & 1.82m+Echelle \\  
2024 11 11 & 17:30:17  &   2460626.22699 &   72499   & 900 & -28.0 & synthe  & 1.82m+Echelle \\  
2024 11 11 & 17:43:39  &   2460626.23627 &   72500   & 900 & -11.8 & synthe  & 1.82m+Echelle \\  
2024 11 12 & 17:11:26  &   2460627.21393 &   72574   & 900 &  14.0 & synthe  & 1.82m+Echelle \\  
2024 11 12 & 17:27:45  &   2460627.22525 &   72575   & 900 &  30.5 & synthe  & 1.82m+Echelle \\  
2024 11 12 & 17:43:59  &   2460627.23652 &   72576   & 900 &  68.0 & synthe  & 1.82m+Echelle \\  
2024 11 14 & 17:06:58  &   2460629.21085 &  123687   & 600 &  73.4 & HR 8634 & 1.22m+B\&C    \\  
2024 11 14 & 17:17:45  &   2460629.21834 &  123688   & 600 &  61.1 & HR 8634 & 1.22m+B\&C    \\  
2024 11 14 & 17:28:08  &   2460629.22555 &  123689   & 600 &  36.1 & HR 8634 & 1.22m+B\&C    \\  
&& \\
\hline
\label{tab:logspectra}
\end{tabular}
\end{table*}

\subsubsection{Archival Radial Velocity Observations and Preliminary RV Fits}
\label{subsubsec:RVcurve}
\begin{figure}
\includegraphics[width=0.99\columnwidth]{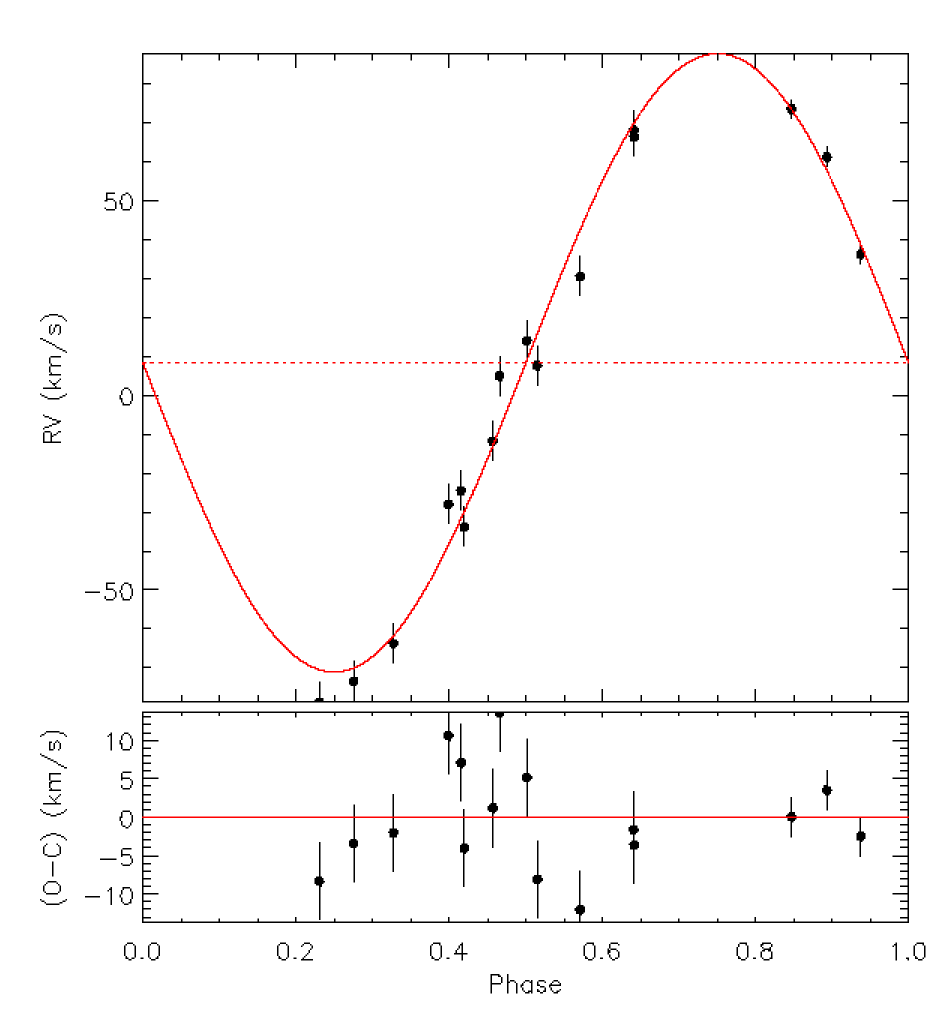}
    \caption{The radial velocity data we acquired in the Asiago Observatory (black data points with error bars) the best fit we achieved with the {\sc rvfit} code (upper panel), and the residuals from the fit (lower panel).}
    \label{fig:dd_crb_asiago_RV}
\end{figure}

We also obtained the published RVs of the system from \citet{for2010} and \citet{blomberg2024}. We also converted the timings of these observations to BJD$_{\rm TDB}$. In order to combine archival RV data with our measurements for further analysis, we needed to determine the systemic velocities (V$_\gamma$), orbital periods (P$_{\rm orb}$), times of periastron passage (T$_{\rm p}$), and semi-amplitudes of the primary component's RVs (K$_1$) for each data set, we used the {\sc rvfit} code \citep{rvfit2015}. The individual RV curves were fitted first separately by adjusting these parameters while keeping the eccentricity (e) and the argument of periastron ($\omega$) fixed at 0, a reasonable assumption given the short orbital period and the occurrence of the secondary eclipses at phase 0.5. The results of the fits are provided in Table~\ref{tab:rvfit}, and the fit to our Asiago Observatory observations is shown in Figure~\ref{fig:dd_crb_asiago_RV}.

\begin{table*}
\caption{Results of the radial velocity fits with the {\sc rvfit} code \citep{rvfit2015}.}
\begin{tabular}{lccc}
\hline
\hline
Parameter			& This Study & \citet{for2010} & \citet{blomberg2024}\\
\hline
\multicolumn{4}{c}{Adjusted Quantities}\\
\hline
$P$ (days)		&0.161748 $\pm$ 0.000002 & 0.161771 $\pm$ 0.000001 & 0.163701 $\pm$ 0.000001\\
$T_p$ (BJD$_{\rm TDB}$)		&2460573.270800 $\pm$ 0.000157 & 2454524.019880 $\pm$ 0.000070 & 2459000.052720 $\pm$ 0.000921\\
$\gamma$ (km/s)	&8.32 $\pm$ 0.98 & -3.87 $\pm$ 0.70 & 3.88 $\pm$ 4.41\\
$K_1$ (km/s)		&79.34 $\pm$ 1.74 & 75.00 $\pm$ 0.80 & 75.00 $\pm$ 3.62\\
\hline
\multicolumn{4}{c}{Derived Quantities}\\
\hline
$a_1\sin i$ ($10^6$ km)	&0.176468 $\pm$ 0.003869 & 0.166838 $\pm$ 0.001770 & 0.168829 $\pm$ 0.075680\\
$f(m_1,m_2)$ ($M_\odot$)	& 0.007071 $\pm$ 0.000225 & 0.00837 $\pm$ 0.00055 & 0.007155 $\pm$ 0.009622\\
\hline
\multicolumn{4}{c}{Other Quantities}\\
\hline
$\chi^2$		&30.91 & 9.69 & 52.38\\
$N_{obs}$ (primary)	&16 & 8 & 38\\
Time span (days)	&55.91745 & 390.94604 & 288.42410\\
$rms_1$ (km/s)	&6.72 & 10.01 & 5.11\\
\hline
\label{tab:rvfit}
\end{tabular}
\end{table*}

After shifting the individual RV curves and computing orbital phases using the orbital periods (P$_{\rm orb}$) and times of periastron passage (T$_{\rm p}$) obtained from the {\sc rvfit} fits, we combined the curves to construct the full radial velocity curve shown in Figure~\ref{fig:DD_CrB_RVs}. This combined RV curve is analyzed together with the light curves in multiple passbands using the second version of the {\sc phoebe} code.

\section{Fundamental Physical and Orbital Parameters}
\label{sec:parameters}

\subsection{Mass Ratio From R{\o}mer Delay}
\label{subsec:mass_ratio}
Since DD~CrB is a single-lined eclipsing binary, the mass ratio (q$ = M_2/M_1$) cannot be determined directly from the radial velocity amplitudes. Consequently, both \citet{for2010} and \citet{lee2017} employed the so-called “q-search” technique, which identifies the mass ratio that minimizes the differences between a model (with all other parameters fixed) and the observations, quantified as $\Sigma(O-C)^2$. However, this approach does not account for degeneracies inherent in light curve modeling, particularly the correlation between mass ratio and orbital inclination. To overcome this limitation, we utilized the R{\o}mer delay arising from the differing distances of the primary component to the observer during the primary and secondary eclipses when the mass ratio is significantly less than unity \citep{kaplan2010}. By measuring the R{\o}mer delay, which shifts the timings of the primary and secondary minima in opposite directions relative to a linear ephemeris, it is possible to constrain the mass ratio. This technique was successfully applied to the prototype of HW~Vir-like binaries by \citet{baran2018}.

We measured both primary and secondary mid-eclipse timings from TESS light curves in sectors 24, 50, and 51. From these measurements, we determined the offset of the secondary minima from orbital phase 0.5, i.e., the expected time of secondary eclipse in the absence of the R{\o}mer delay by using the simple relation $\Delta t = t_s - t_p - P_{orb}/2$ given by \citet{kaplan2010}. To avoid potential biases from variations in the orbital period, we employed the linear ephemeris elements calculated for each TESS sector during the computation of the R{\o}mer delay. We found $\Delta t = 2.55 \pm 0.11$~s, which agrees with the theoretical value of $2.7 \pm 1.4$~s derived by \citet{lee2017} based on their orbital period and radial velocity amplitude. Although the timing uncertainties of individual secondary minima are relatively large ($\sim 8$~s), the continuous TESS coverage provides $\sim 100$ minima of each type (primary and secondary) for DD~CrB, significantly improving the precision of the $\Delta t$ measurement. 

It should be noted that the method of \citet{kwee1956} used to measure mid-eclipse timings tends to underestimate the uncertainties. Nevertheless, linear fits to the minima timings in each sector yield reduced $\chi^2_\nu$ values very close to 1.0, so no additional scaling of the TESS timing uncertainties was necessary. The $\Delta t = 2.55 \pm 0.11$\,s value we find gives a mass ratio of $q = 0.299 \pm 0.009$ using the parameters (P$_{\rm orb}$,K$_1$) from the light and radial velocity curves, and Eq. \ref{eq:mass_ratio} assuming a circular orbit \citet{baran2018}.

\begin{equation}
  q = (\frac{\pi~c~\Delta t}{P_{orb}~K_1} +1)^{-1}
  \label{eq:mass_ratio}
\end{equation}

Then it is possible to derive the mass of the sdB primary companion following \citet{kaplan2010} based on Eq. \ref{eq:mass_pri} as $M_1 = 0.417 \pm 0.032~M_{\odot}$,
\begin{equation}
  M_1 = \frac{\pi^2~c^3}{2~G~P_{orb}^2} \frac{(1 + q)^2}{(1 - q)^3} \Delta t^3
  \label{eq:mass_pri}
\end{equation}
and that of the secondary from the mass ratio to be $M_2 = 0.125 \pm 0.006~M_{\odot}$.

\subsection{Analyses of the Light \& Radial Velocity Curves}
\label{subsec:lc_rv_analysis}

\begin{figure*}
	\includegraphics[width=0.85\paperwidth]{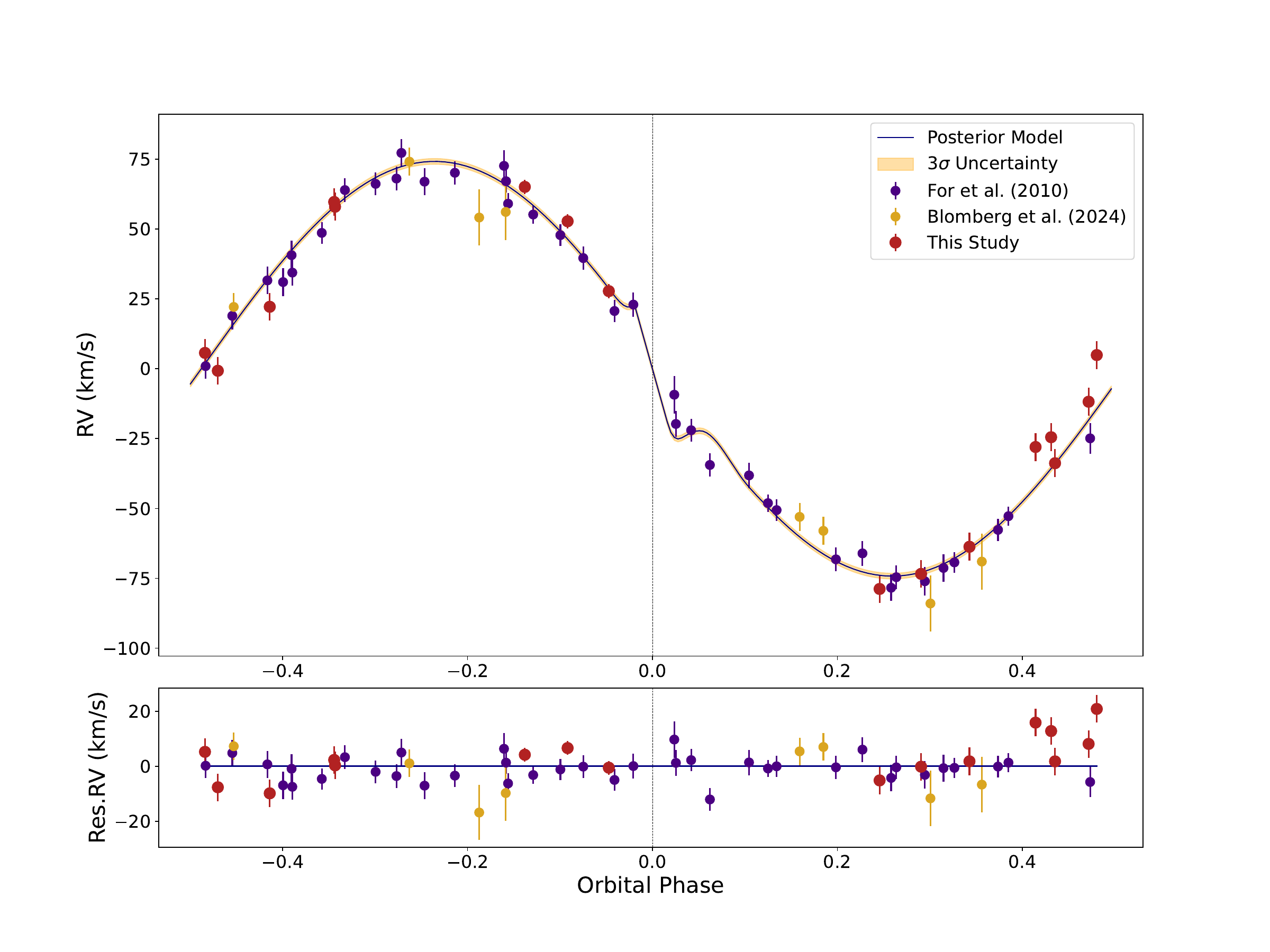}
    \caption{Radial velocities of DD~CrB obtained with different spectrographs (data points) and their model achieved using {\sc phoebe}. Semi-amplitude of the radial velocity curve is found out to be K$_{1} = 73.40^{0.76}_{0.77}$ km/s from the {\sc phoebe} model.} 
    \label{fig:DD_CrB_RVs}
\end{figure*}

We analyzed the multi-color light curves obtained using SDSS filters and the radial velocity curves of DD~CrB using version 2.17 of the {\sc phoebe} software package \citep{conroy2020,conroy2021}. Primary minima were found to be shallower in the TESS light curves due to a potential contamination source in the $21\times21$ arcsec-pixels of TESS although there is no indication of such a source in Gaia observations. For computational efficiency, we first modelled the light curves in the SDSS-$u' g' r' i' z'$ passbands separately, using the combined RV curve described in Sec.~\ref{subsubsec:RVcurve}. To account for the variable orbital period, we shifted and scaled the RV curve in Fig.~\ref{fig:DD_CrB_RVs} along the time axis to match the epoch of the light curve in each passband, using the super-conjunction time (T$_0$, the time of primary minimum) and the orbital period (P$_{\rm orb}$) derived from that light curve. This procedure reduces the number of free parameters during modeling by avoiding the need to fit the orbital period derivative ($\dot{P}$).

\begingroup
\renewcommand{\arraystretch}{1.25} 
\begin{table*}
\centering
\caption{Median values and asymmetric uncertainties corresponding to the 16th and 84th percentiles for the physical and orbital parameters of DD~CrB derived from our analysis.} 

\begin{adjustbox}{width=\textwidth}
        \begin{tabular}{llccccc>{\bfseries\boldmath}c}
 \hline
  \multicolumn{8}{c}{Light \& Radial Velocity Curve Models} \\
 \hline
 Symbol & Parameter (Unit) & & & Values &  \\
\hline
\multicolumn{2}{l}{Passband:} & SDSS-u$^\prime$ & SDSS-g$^\prime$ & SDSS-r$^\prime$ & SDSS-i$^\prime$ & SDSS-z$^\prime$ & Simultaneous\\
\hline
$i$&Orbital inclination (deg) &$86.372^{+0.099}_{-0.100}$&$86.812^{+0.037}_{-0.038}$&$86.487^{+0.038}_{-0.036}$&$86.396^{+0.027}_{-0.031}$&$86.740^{+0.100}_{-0.100}$&$85.814^{+0.013}_{-0.011}$\\
$q$&Mass ratio (-) 
&$0.3002^{+0.0033}_{-0.0035}$&$0.2988^{+0.0037}_{-0.0039}$&$0.3008^{+0.0030}_{-0.0035}$&$0.3009^{+0.0039}_{-0.0039}$&$0.2997^{+0.0033}_{-0.0034}$&$0.3042^{+0.0050}_{-0.0017}$\\
R$_1$&Radius of the primary (R$_{\odot}$) &$0.1668^{+0.0011}_{-0.0010}$&$0.1695^{+0.0003}_{-0.0004}$&$0.1691^{+0.0003}_{-0.0004}$&$0.1683^{+0.0003}_{-0.0004}$&$0.1709^{+0.0010}_{-0.0011}$&$0.1644^{+0.0002}_{-0.0002}$\\
R$_2$&Radius of the secondary (R$_{\odot}$) &$0.1572^{+0.0006}_{-0.0006}$&$0.1542^{+0.0003}_{-0.0002}$&$0.1574^{+0.0003}_{-0.0003}$&$0.1579^{+0.0003}_{-0.0003}$&$0.1553^{+0.0008}_{-0.0008}$&$0.1619^{+0.0002}_{-0.0002}$\\
T$_{2}$&Surface temperature of the primary (K) &$29230$~(fixed)&$29230$~(fixed)&$29230$(fixed)&$29230$(fixed)&$29230$(fixed)&$29230$(fixed)\\
T$_{2}$&Surface temperature of the secondary (K) &$3090^{+210}_{-210}$&$2980^{+200}_{-200}$&$2920^{+160}_{-170}$&$2790^{+150}_{-160}$&$3000^{+200}_{-200}$&$2357^{+70}_{-69}$\\
A$_1$&Albedo of the primary (-) &$1.0000$~(fixed)&$1.0000$~(fixed)&$1.0000$~(fixed)&$1.0000$~(fixed)&$1.0000$~(fixed)&$1.0000$~(fixed)\\
A$_2$&Albedo of the secondary (-) &$1.0000$~(fixed)&$1.1560^{+0.0079}_{-0.0079}$&$1.3290^{+0.0120}_{-0.0120}$&$1.4390^{+0.0120}_{-0.0120}$&$1.6520^{+0.0340}_{-0.0340}$&$1.0000$~(fixed) \\
g$_1$&Gravity darkening of the primary&$1.00$~(fixed)&$1.00$~(fixed)&$1.00$~(fixed)&$1.00$~(fixed)&$1.00$~(fixed)&$1.00$~(fixed)\\
g$_2$&Gravity darkening of the secondary&$0.32$~(fixed)&$0.32$~(fixed)&$0.32$~(fixed)&$0.32$~(fixed)&$0.32$~(fixed)&$0.32$~(fixed)\\
a&Semi-major axis (R$_{\odot}$) &$1.019 \pm 0.039$&$1.019 \pm 0.039$&$1.019 \pm 0.038$&$1.019 \pm 0.040$&$1.019 \pm 0.040$&$1.020 \pm 0.018$\\
M$_1$&Mass of the primary (M$_{\odot}$) &$0.417 \pm 0.050$&$0.417 \pm 0.050$&$0.417 \pm 0.040$&$0.417 \pm 0.060$&$0.417 \pm 0.050$&$0.417 \pm 0.022$\\
M$_2$&Mass of the secondary (M$_{\odot}$) &$0.126 \pm 0.017$&$0.124 \pm 0.017$&$0.127 \pm 0.017$&$0.127 \pm 0.017$&$0.125 \pm 0.017$&$0.127 \pm 0.017$\\
L$_1$&Luminosity of the primary (L$_{\odot}$) &$18.524 \pm 0.561$&$19.129 \pm 0.526$&$19.039 \pm 0.524$&$19.859 \pm 0.519$&$19.446 \pm 0.577$&$17.989 \pm 0.494$\\
L$_2$&Luminosity of the secondary (L$_{\odot}$) &$0.0020 \pm 0.0006$&$0.0017 \pm 0.0005$&$0.0016 \pm 0.0004$&$0.0014 \pm 0.0003$&$0.0018 \pm 0.0005$&$0.0007 \pm 0.0001$\\
log~g$_1$&Surface gravity of the primary (cgs) &$5.614 \pm 0.052$&$5.600 \pm 0.052$&$5.602 \pm 0.042$&$5.606 \pm 0.063$&$5.593 \pm 0.052$&$5.627 \pm 0.023$\\
log~g$_2$&Surface gravity of the secondary (cgs) &$5.146 \pm 0.059$&$5.155 \pm 0.060$&$5.148 \pm 0.058$&$5.145 \pm 0.058$&$5.153 \pm 0.059$&$5.123 \pm 0.058$\\
 \hline
 \multicolumn{2}{l}{Limb Darkening Coefficients:}\\
 \hline
 $s_{1}, s_{2}$&Square Root Law Coeffs. (primary) &$-0.0765,-0.0765$&$-0.1037,0.5640$&$-0.0942,0.4777$&$-0.0872,0.4174$&$-0.0928,0.3831$&-\\
 $q_{1}, q_{2}$&Quadratic Law Coeffs. (secondary) &$0.4839,0.2728$&$0.4274,0.3381$&$0.4151,0.3298$&$-0.0071,0.8057$&$0.1954,0.3722$&-\\
\hline
\label{tab:modelparams}
        \end{tabular}
    \end{adjustbox}
\end{table*}
\endgroup

We fixed the effective temperature of the primary companion to the spectroscopic value determined by \citet{for2010}, T$_1 = 29{,}230$~K, while the temperature of the secondary (T$_2$) was treated as a free parameter. The mass ratio was incorporated as a Gaussian prior centered on the value derived from the R{\o}mer delay, with a standard deviation equal to its uncertainty. For the other adjusted parameters—orbital inclination (i$ = 86.5 \pm 0.5^\circ$), equivalent radii of the primary and secondary (R$_1 = 0.165 \pm 0.015~R_\odot$, R$_2 = 0.155 \pm 0.015~R_\odot$), and secondary temperature (T$_2 = 3000 \pm 300$~K)—we adopted Gaussian priors centered on values consistent with both \citet{for2010} and \citet{lee2017}. The standard deviations of these priors were deliberately set larger than the uncertainties reported in the literature to account for the well-known underestimation of parameter errors by the Wilson-Devinney code \citep{wilson_devinney1971} and the {\sc simplex} optimization algorithm used in those studies. Since DD~CrB is a single-lined eclipsing binary, we inverted the usual constraint in {\sc phoebe}, solving for the semi-major axis ($a$) instead of the primary mass.

We employed the square-root limb darkening law for the primary component, as \citet{claret2020} demonstrated that, for O/B stars, particularly in the UV—the square-root law provides a better fit to model atmospheres than the quadratic law. For the much cooler secondary, we adopted the quadratic law. The relevant coefficients were obtained from the tables of \citet{claret2011}, assuming effective temperatures of 29,000~K and 3,500~K, $\log g = 5.0$, and microturbulent velocities of 1.0 and 2.0~km/s for the primary and secondary, respectively, based on ATLAS models for solar metallicity using the flux conservation method.

Although we attempted interpolations and extrapolations across several tables \citep{claret2000,claret2004,claret2011,claret2020,claret2022}, as the coefficients for the secondary fell outside the tabulated temperature and surface gravity ranges, we ultimately adopted the coefficients corresponding to the nearest atmospheric parameters, following consultation with Antonio Claret (private communication). We also tested various limb darkening laws and coefficient values; these variations did not significantly affect the models, provided the stellar parameters used to derive the coefficients were close to the predicted parameters for the DD~CrB components.

Ultimately, we fixed the limb darkening coefficients at the values listed in Table~\ref{tab:modelparams}, reducing the number of free parameters by four for each band. Bolometric limb darkening coefficients were determined and fixed for the same atmospheric parameters and limb darkening laws following \citet{claret2000}. The bolometric gravity darkening parameters were fixed at g$_1 = 1.00$ for the hot primary with a radiative envelope, and g$_2 = 0.32$ for the cooler secondary with a convective envelope.

The light curves of HW~Vir-like systems, close binaries with a subdwarf B-type primary and a cool M-dwarf secondary, are dominated by a strong reflection effect, which increases toward longer wavelengths. Consequently, many studies \citep[e.g.,][]{for2010,lee2017,esmer2021,esmer2022}, 
including those focused on DD~CrB, have assumed bolometric albedos of 1.00 for the primary and even greater than unity for the secondary, despite the lack of physical justification. This is partly due to the limited treatment of irradiation effects in existing models. The day sides of the cooler companions are strongly irradiated by the hot subdwarf, but the detailed response of their stellar atmospheres to this external energy is not well understood. Since the reflection effect increases with wavelength, the albedo of the secondary (A$_2$) must be artificially enhanced in bands other than SDSS-$u^\prime$. Accordingly, we fixed A$_2 = 1.00$ in SDSS-u$^\prime$, and adopted Gaussian priors for the other passbands, centered at 1.25, 1.35, 1.45, and 1.60 for SDSS-g$^\prime$, r$^\prime$, i$^\prime$, and z$^\prime$, respectively, with standard deviations of 0.10.

We then attempted simultaneous solutions of all light curves in different passbands together with the radial velocity curve, testing fixed values of the secondary albedo (A$_2 = 0.5, 0.8, 1.0$) as well as allowing it to vary. During these experiments, we observed degeneracies between the albedo, surface temperature, and radius of the secondary component. Satisfactory fits were not obtained for A$_2 = 0.5$ and $0.8$, whereas for A$_2 = 1.0$ we obtained the values listed in the last column of Table~\ref{tab:modelparams}. The radii of both components were slightly larger compared to single-band models, but the most notable difference was found for the surface temperature of the secondary. In the multi-band fit, we obtained T$_2 = 2357^{+70}_{-69}$,K, which is 400–700,K lower than the temperatures derived from single-color light curve analysis. We attribute this difference to the efficiency of the irradiation treatment \citep{horvat2019} implemented in {\sc phoebe}. 

\begin{figure*}
	\includegraphics[width=0.85\paperwidth]{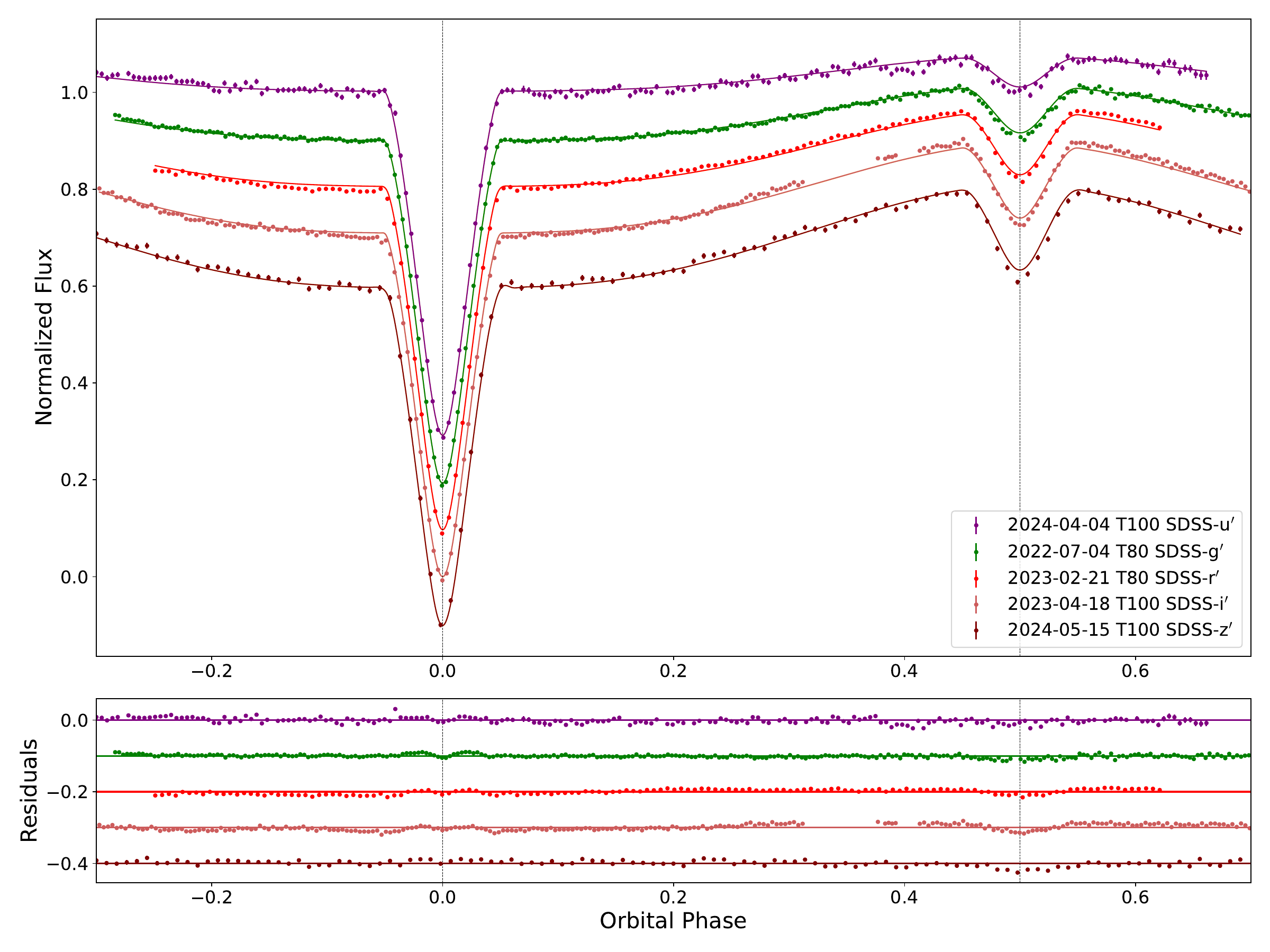}
    \caption{Light curves of DD~CrB in SDSS-ugriz (data points in different colors as given in the legend) and the synthetic light curves (with the same colors) for the model achieved using {\sc phoebe} by simultaneously modelling all the light curves in different passbands with a fixed value of the albedo for both companions (A$_1$ = A$_2 = 1.0$) (upper panel) , and the residuals from the models (lower panel).}
    \label{fig:lcmodel_phoebe_mcmc_all}
\end{figure*}

Although satisfactory fits were not achieved with lower A$_2$ values, the effective temperature of the secondary increased to $\sim3000$,K for A$_2 = 0.5$ and to $\sim2800$,K for A$_2 = 0.8$. This trend is consistent with the single-band models: for SDSS-u$^\prime$, the surface temperature was $3090 \pm 210$,K with A$_2 = 1.0$, decreasing systematically toward longer wavelengths, where the adjusted A$_2$ values increased to 1.156, 1.329, and 1.439 for SDSS-g$^\prime$, r$^\prime$, and i$^\prime$, respectively. Although A$_2 = 1.652$ was found for SDSS-z$^\prime$, the radii of both components were found to be smaller. These results confirm the expected degeneracy between albedo, surface temperature, and stellar radii in modeling the reflection effect.

We employed 16 walkers and 5000 steps, discarding the first 1000 steps as burn-in, to sample the prior distributions and obtain posterior distributions based on the likelihood of the model given these priors for all models presented here. Our results are reported in Table~\ref{tab:modelparams}, and the corner plots for SDSS-$u^\prime$, including both the simultaneous solutions obtained by fixing $A_2$ to 1.0 and those with $A_2$ treated as a free parameter, are provided in Appendix \ref{app:cornerplots}. Although the posterior distributions of some fit parameters are less well-behaved for the model in which A$_2$ was fixed to 1.0, we adopt this model as the most physically meaningful, given the system’s properties and the fact that A$_2$ loses its physical interpretation when exceeding unity. Although we paid utmost attention to the underestimation of uncertainties in derived parameter values and followed a probabilistic modelling scheme, we could not avoid the well-known problem of formal uncertainties in the Wilson-Devinney technique as inherited by {\sc phoebe}, especially considering the unrealistically small uncertainties in the radii and luminosity values we derived for the secondary component. This is also due to the fast convergence of the parameter values, resulting in very tight distributions from which the posteriors are sampled. In all cases, because the orbital size, mass ratio, and thus the component masses are constrained by the radial velocity observations and the measured R{\o}mer delay, we were able to reliably derive the parameters required for our eclipse timing variation analysis. We provide the fits, obtained using the median values of the fitted parameters, superimposed on observational in all SDSS filters and their residuals in Figure \ref{fig:lcmodel_phoebe_mcmc_all}.

\section{Eclipse Timing Analysis}
\label{sec:etv_analysis}

To investigate the eclipse timing variations (ETVs) of DD~CrB, we compiled eclipse timings from the literature, archival data, and our own new measurements. Using the reference linear ephemeris provided by \citet{wolf2021}, we constructed the ETV diagram using this linear ephemeris as:
\begin{equation}
\mathrm{Min\,I} (\mathrm{BJD}_{\mathrm{TDB}}) = 2455611.92657 + 0.161770446 \times E
\label{eq:ephemeris_wolf}
\end{equation}
Two measurements show isolated deviations from the general behaviour of the ETV diagram and are inconsistent with neighbouring epochs. Their deviations lie well beyond the $3\sigma$ scatter of the remaining data, and we therefore exclude them from the analysis as likely non-astrophysical in origin. The final data set used in the ETV analysis consists of 618 precise primary eclipse timings.

\begin{figure}
	\includegraphics[width=\columnwidth]{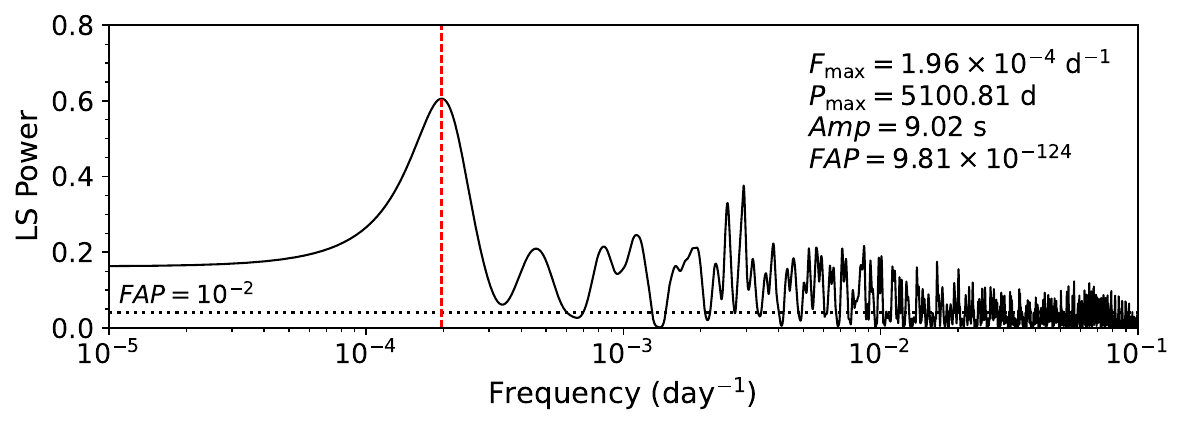}
    \caption{Lomb-Scargle power spectrum of the DD~CrB eclipse timing residuals. The red dashed vertical line marks the frequency with the maximum power ($f_{\rm max} \approx 1.96 \times 10^{-4}$ d$^{-1}$). The horizontal dotted line represents the False Alarm Probability (FAP) level of $10^{-2}$, highlighting the high significance of the detected signal.}
    \label{fig:ls_spectrum}
\end{figure}

To characterize the cyclic trend, which appears to span nearly a complete cycle within our observational baseline, we performed a Lomb–Scargle (LS) frequency analysis \citep{lomb1976, scargle1982} on the full timing dataset, using the implementation in \textsc{Astropy} \citep{astropy2013, astropy2018, astropy2022}. The resulting periodogram (Fig.~\ref{fig:ls_spectrum}) shows a well-defined peak at $f_{\rm max} = 1.96 \times 10^{-4}$~d$^{-1}$, corresponding to a period of $P_{\rm max} \approx 5101$~d ($\sim 14$ yr). The amplitude inferred from the LS model is about $9$ s. The signal is statistically significant, with a false-alarm probability of $9.8 \times 10^{-124}$.

To explain this periodicity, we modeled the data assuming a Keplerian orbit. We tested two families of models: (i) a linear ephemeris combined with the light–time effect (LiTE), and (ii) a linear ephemeris combined with LiTE and a quadratic (parabolic) term. For each model family, the orbital eccentricity of the proposed third body was either fixed at zero (circular orbit) or treated as a free parameter; in the circular-orbit case ($e=0$), the argument of periastron was fixed to $\omega = 0$, following the formulation in \citet{esmer2021}.

\begin{figure*}
  \centering
  \includegraphics[width=0.85\paperwidth]{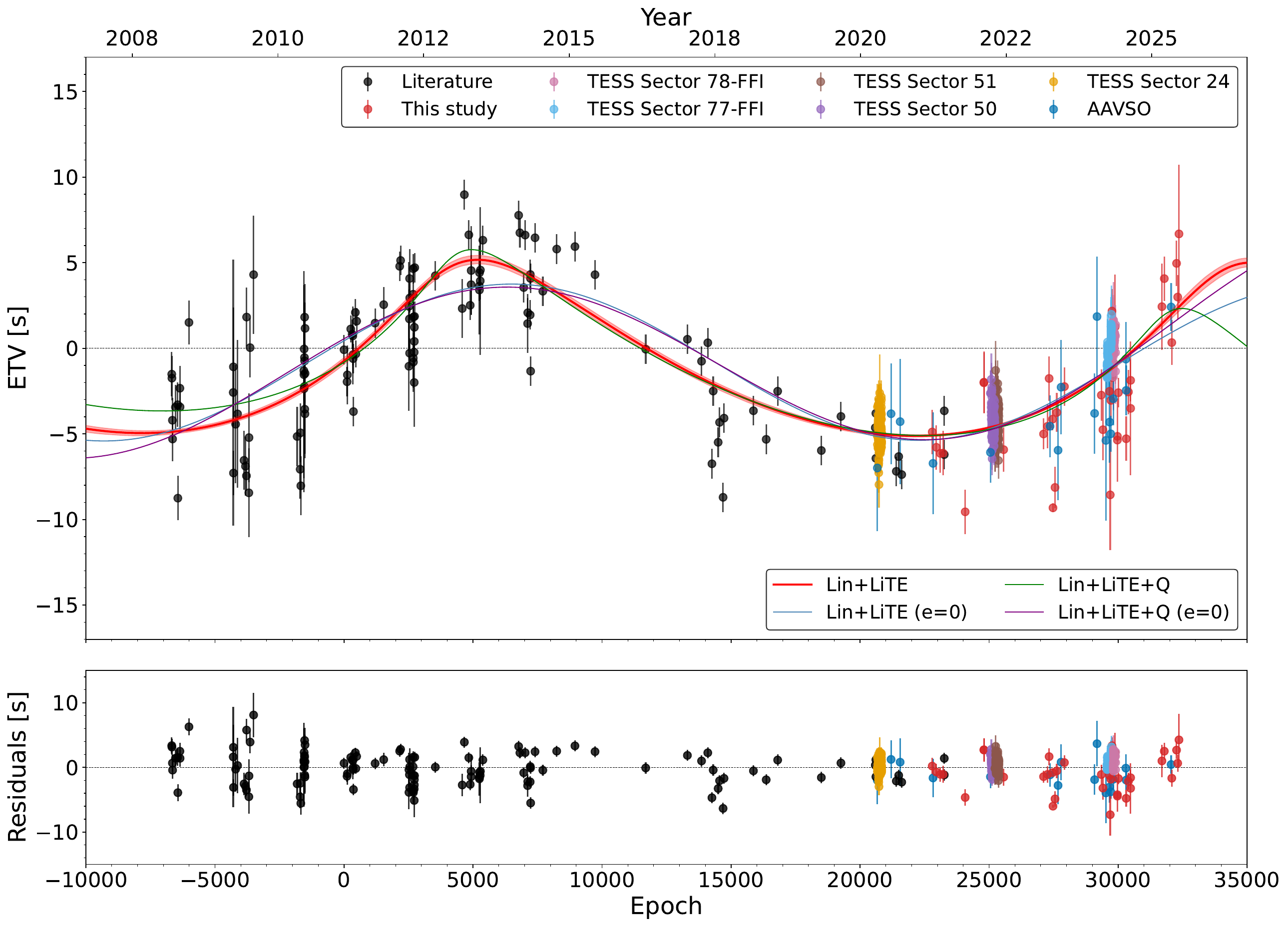}
  \caption{ETV diagram of the DD~CrB system with four different ETV models overlaid. The curves correspond to the Linear+LiTE (red), Linear+LiTE ($e=0$) (blue), Linear+Quadratic+LiTE (green), and Linear+Quadratic+LiTE ($e=0$) (orange) models based on the median values of the fit parameters. The shaded region in red indicates the $1\sigma$ uncertainty band of the preferred model. The residuals are shown with respect to the preferred Linear+LiTE model.}
  \label{fig:DDCrB_All_model}
\end{figure*}

We sampled the model parameters using a Markov chain Monte Carlo (MCMC) method to explore their posterior probability distributions and to derive parameter uncertainties. We employed the \textsc{emcee} package \citep{foremanmackey2013}, adopting Gaussian prior distributions centered on the parameter values obtained from a preliminary non-linear least-squares fit carried out using the \textsc{lmfit} package \citep{lmfit2016}, with the corresponding uncertainties used as prior widths.The sampling was performed with 500 walkers evolved for 6000 steps each. The first 1000 steps of each chain were discarded as burn-in. We compared the models using the reduced chi-squared ($\chi^2_\nu$), the Akaike Information Criterion (AIC), and the Bayesian Information Criterion (BIC).

The results of the model comparison are summarised in Table~\ref{tab:lite_four_models}. Among the four tested configurations, we find that the linear ephemeris combined with the light-time effect (LiTE), in which the orbital eccentricity of the third body is treated as a free parameter, provides the most appropriate description of the data. Although the Linear+Quadratic+LiTE model yields a slightly lower reduced chi-squared value, the improvement is marginal and is accompanied by an increase in model complexity due to the additional quadratic term.

In contrast, the linear+LiTE model achieves a comparable goodness of fit while remaining statistically more parsimonious. We therefore select this model as the preferred solution and indicate it with boldface in Table~\ref{tab:lite_four_models}. Furthermore, we find that both models allowing for a non-zero eccentricity of the tertiary orbit perform significantly better than their circular counterparts, indicating that an eccentric outer orbit is required to reproduce the observed eclipse timing variations. The corner plot of the posterior distributions for the Linear+LiTE model is presented in Appendix \ref{app:cornerplots}.

Based on the linear part of the preferred Linear+LiTE model (using dP and dT values in Table-\ref{tab:lite_four_models} for the preferred model), linear ephemeris in Eq. \ref{eq:ephemeris_wolf} is updated as
\begin{align}
\mathrm{Min\,I}~(\mathrm{BJD}_{\mathrm{TDB}}) =
(2455611.9265712 \pm 1.7\times10^{-6}) \nonumber\\
+ (0.16177044594 \pm 5\times10^{-11})\,E
\end{align}

\begin{table*}
  \centering
  \footnotesize
  \renewcommand{\arraystretch}{1.3}
  \caption{Posterior median values and uncertainties for the parameters of four ETV models for DD~CrB obtained via MCMC sampling. The preferred model is highlighted in bold. Units are indicated in parentheses. dP and dT provide the corrections for the linear ephemeris based on the linear model, while T$_{\rm p,3}$ and P$_{\rm 3}$ are the periastron passage and the period of the third body, and A$_{\rm 3}$ is the amplitude of the LiTE model.}
  \label{tab:lite_four_models}
  \resizebox{\textwidth}{!}{
    \begin{tabular}{l c c >{\bfseries\boldmath}c c}
      \hline
      \multicolumn{5}{c}{\textbf{ETV Models}} \\
      \hline
      Parameter (Unit) & Linear+Quad+LiTE & Linear+Quad+LiTE ($e=0$) & \textbf{Linear+LiTE} & Linear+LiTE ($e=0$) \\
      \hline
      $dP$ (day) & $-3.25 \times 10^{-12} \pm 4.95 \times 10^{-11}$ & $-9.12 \times 10^{-13} \pm 4.97 \times 10^{-11}$ & $-6.14 \times 10^{-11} \pm 4.83 \times 10^{-11}$ & $1.45 \times 10^{-11} \pm 4.88 \times 10^{-11}$ \\
      $dT$ (day) & $1.35 \times 10^{-5} \pm 2.72 \times 10^{-6}$ & $-1.90 \times 10^{-5} \pm 5.35 \times 10^{-6}$ & $1.17 \times 10^{-6} \pm 1.65 \times 10^{-6}$ & $-9.61 \times 10^{-6} \pm 1.09 \times 10^{-6}$ \\
      $T_{\mathrm{p,3}}$ (BJD) & $2460766.0 \pm 59.3$ & $2460748.2 \pm 139.5$ & $2461109.3 \pm 74.3$ & $2460487.8 \pm 20.3$ \\
      $P_{\mathrm{3}}$ (day) & $4474.0 \pm 59.9$ & $5526.9 \pm 227.2$ & $4852.3 \pm 35.9$ & $5104.1 \pm 41.5$ \\
      $A_{\mathrm{3}}$ (s) & $4.68 \pm 0.15$ & $5.11 \pm 0.34$ & $5.10 \pm 0.13$ & $4.57 \pm 0.08$ \\
      $e$ (-) & $0.62 \pm 0.05$ & 0 (fixed) & $0.46 \pm 0.04$ & 0 (fixed) \\
      $\omega$ (deg) & $59.3 \pm 4.1$ & 0 (fixed) & $61.0 \pm 5.3$ & 0 (fixed) \\
      $a_{12}\sin i$ (AU) & $0.0099 \pm 0.0003$ & $0.0102 \pm 0.0007$ & $0.0105 \pm 0.0003$ & $0.0092 \pm 0.0002$ \\
      $f(m_3)$ (M$_\odot$) & $6.43 \times 10^{-9} \pm 7.01 \times 10^{-10}$ & $4.68 \times 10^{-9} \pm 1.01 \times 10^{-9}$ & $6.52 \times 10^{-9} \pm 5.37 \times 10^{-10}$ & $3.94 \times 10^{-9} \pm 2.27 \times 10^{-10}$ \\
      $m_3\sin i$ (M$_{\rm Jup}$) & $1.30 \pm 0.06$ & $1.17 \pm 0.09$ & $1.30 \pm 0.06$ & $1.10 \pm 0.04$ \\
      $a_3\sin i$ (AU) & $4.34 \pm 0.08$ & $4.99 \pm 0.16$ & $4.58 \pm 0.08$ & $4.74 \pm 0.08$ \\
      $Q$ (day\,cycle$^{-2}$) & $-3.83 \times 10^{-14} \pm 6.63 \times 10^{-15}$ & $3.12 \times 10^{-14} \pm 1.67 \times 10^{-14}$ & --- & --- \\
      \hline
      $\chi^2$ & 1981.70 & 2150.19 & 2003.08 & 2153.79 \\
      $\chi^2_\nu$ & 3.2014 & 3.4625 & 3.2308 & 3.4627 \\
      AIC & 1997.70 & 2162.19 & 2017.08 & 2163.79 \\
      BIC & 2033.22 & 2188.84 & 2048.17 & 2185.99 \\
      \hline
    \end{tabular}
  }
\end{table*}

Previous O–C analyses of the system suggested that the light-time effect (LiTE) is the most plausible explanation for the observed periodic variations in the orbital period \citep{wolf2021}. In those studies, the orbital eccentricity ($e$) of the proposed third body was fixed at zero. While our findings agree with the LiTE explanation, with the extended observational baseline available in our work, we treated the eccentricity as a free parameter and found it to be 0.46.

To evaluate whether the observed cyclic modulation in DD~CrB can be attributed to magnetic activity cycles in the cool secondary star as suggested as a possibility by \citet{pulley2025}, we computed the energy requirements of the Applegate mechanism \citep{applegate1992}. For this calculation, we employed the three analytical formulations described by \citet{volschow2016}: the thin shell approximation, the constant density model, and the two zone model. We calculated the ratio of the required energy ($\Delta E$) to the total energy available from the secondary ($E_{\mathrm{sec}}$) for each formulation. The resulting ratios are $\Delta E / E_{\mathrm{sec}} \approx 0.6$ for the thin shell approximation, $\Delta E / E_{\mathrm{sec}} \approx 7.1 \times 10^{2}$ for the constant density model, and $\Delta E / E_{\mathrm{sec}} \approx 11.3$ for the two zone model. For the Applegate mechanism to be physically viable, the condition $\Delta E \ll E_{\mathrm{sec}}$ must be satisfied \citep{volschow2016}. It has been demonstrated that the thin-shell approximation systematically underestimates the required energy, while the constant density model tends to overestimate it. The thin-shell value formally lies below unity but still demands that a substantial fraction of the star's luminosity be diverted into magnetic restructuring, whereas the constant density estimate exceeds the available energy by several orders of magnitude and is clearly unphysical. The two-zone model provides the most accurate analytic approximation, reproducing detailed numerical results within a deviation of less than 25 percent \citep{volschow2016}. Since this robust formulation yields an energy requirement of $\Delta E / E_{\mathrm{sec}} \approx 11.3$, which exceeds the total available energy of the secondary by a wide margin, we conclude that the Applegate mechanism is energetically implausible and cannot explain the observed ETV behavior of DD~CrB.

Following the evaluation of the Applegate mechanism, we also tested the alternative spin-orbit coupling model proposed by \citet{Lanza2020}. In this framework, the orbital period modulation is driven by a cyclic exchange of angular momentum between the spin of the magnetically active component and the orbital motion, mediated by a long-lived non-axisymmetric internal magnetic field, rather than by the structural deformation of the star.

We applied the energetic criterion of \citet{Lanza2020} to the secondary component of DD~CrB by computing the variation in the rotational kinetic energy, $\Delta \mathcal{T}$, required to sustain the observed orbital period modulation and comparing it to the energy that can be supplied by the stellar luminosity, $L$, over a modulation cycle. Our analysis yields a characteristic energy replenishment timescale of $\Delta \mathcal{T}/L \approx 118.3\,\mathrm{yr}$, whereas the observed modulation period is $P_{\mathrm{mod}} \approx 13.3\,\mathrm{yr}$. Since $\Delta \mathcal{T}/L$ exceeds $P_{\mathrm{mod}}$ by a factor of approximately $8.9$, the secondary star is unable to replenish the required energy within a single modulation cycle. This result indicates that the spin-orbit coupling mechanism is energetically implausible for DD~CrB. We therefore conclude that magnetic activity cycles, whether operating through the classical Applegate mechanism or through the spin-orbit coupling model of Lanza, cannot account for the cyclic eclipse timing variations observed in DD~CrB. 

The failure of magnetic activity driven mechanisms implies that the observed cyclic modulation in the ETV diagram of DD~CrB cannot be explained by intrinsic processes associated with the secondary star. We therefore also assessed whether apsidal motion is present in the eclipsing binary. Apsidal motion requires an eccentric inner orbit and produces anti-correlated timing variations between the primary and secondary eclipses. In DD~CrB, neither condition is satisfied, as the primary and secondary eclipse timings do not exhibit opposite phase behavior and the orbit of the eclipsing binary is consistent with being circular.

Having found no physically viable explanation based on magnetic activity cycles or apsidal motion, we interpret the observed cyclic ETVs as arising from a Keplerian modulation of the binary orbit caused by the presence of an additional gravitationally bound companion.

Our best-fit solution for the LiTE indicates that the third body follows an orbit with a period of P$_3 = 4852.3 \pm 35.9$ days and an eccentricity of e$_3 = 0.46 \pm 0.04$. The minimum mass of this companion is m$_3 \sin i = 1.30 \pm 0.06~M_\mathrm{Jup}$, in agreement with the results reported by \citet{wolf2021}. The projected orbit of the third body has a semi-major axis of a$_3 \sin i = 4.58 \pm 0.08$ au, while the DD~CrB binary revolves around the common center of mass with a projected semi-major axis of a$_{12} \sin i = 0.0105 \pm 0.0003$ au.

The physical parameters required for the light curve, radial velocity, and ETV modeling of DD~CrB were obtained from the CuPS-ETV (Circumbinary Planet System Catalog)\footnote{\url{https://cups.astrotux.org}}. For a detailed description of the catalog and its methodology, see \citet{2025CoSka..55c.301S}.

\section{Discussion}
\label{sec:discussion}
DD~CrB is an intriguing system, consisting of a B-type subdwarf and an M-type main-sequence star in a close orbit whose orbital period varies with time. Since almost a full cycle of this periodic change is now apparent in the O–C diagram, we aimed to investigate its origin. To this end, we modeled the multi-color light curves obtained from two observatories in T\"urkiye, combined them with archival and recent radial velocity observations collected with two telescopes at the Asiago Observatory in Italy, and analyzed the R{\o}mer delay between the primary and secondary minima, which allowed us to constrain the mass ratio of the system.

We first modeled the multi-color light curves of the system together with the radial velocity observations, initially for each light curve separately and then simultaneously. The limb- and gravity-darkening coefficients were fixed according to the stellar physical parameters. We observed degeneracies between the albedo, surface temperature, and radius of the secondary (cooler) component, which cannot be directly constrained by either photometry or spectroscopy due to its faintness. Since this degeneracy cannot be resolved with the current data, we adopted the maximum physically plausible albedo of 1.0, which yielded a surface temperature of $2357$ K, significantly lower than previous literature values 
($3100$ K \citep{for2010}, $3089$ K \citep{lee2017}). Nonetheless, they agree with each other to within $1\sigma$ due to the large uncertainty values. The radius of the secondary R$_2 = 0.1619~R_{\odot}$ is also consistent with the published results, even though some studies reported albedo values exceeding unity. We attribute this lower surface temperature to the efficiency of the heat redistribution method implemented in the {\sc phoebe} modelling suite \citep{horvat2019} as well as our lack of understanding of the secondary since it contributes next to nothing to the total light. 
Because the energy received from the hot primary by the cooler secondary is partly reflected and partly absorbed as heat, modeling the redistribution of this energy across the secondary’s surface is crucial. In models with single light curves or simultaneous multi-band light curves using albedos larger than unity, as done in many previous studies, despite being unphysical, we obtained even smaller surface temperatures for the secondary. It should be stressed that we cannot unambiguously determine the surface temperature of the secondary, which we do not see in either photometry or spectroscopy, because it contributes to the total light only by reflection. Therefore, the suggested values for the surface temperature of the secondary, and hence its luminosity, should be considered with caution. In our simultaneous multi-band model, a physically consistent solution was achieved with a single albedo value of 1.0. Although this implies that all incident energy is reflected, the inability to model the light curves with lower albedo values indicates that this solution represents the only physically motivated model with the current observations. 

We then modeled the ETV diagram of the system using the total mass and the relative orbital size derived from the simultaneous light and radial velocity curve analysis with {\sc phoebe}. The mass ratio, constrained via the R{\o}mer delay between primary and secondary minima, was held fixed in all ETV models, ensuring consistency across the analysis. We found that the observed variations in the arrival times of light can be explained by the changing distance between DD~CrB and the observer due to its orbital motion around a common center of mass with a tertiary body of $1.30 \pm 0.06~M_{\rm Jup}$ on an eccentric orbit (e$_3 = 0.46 \pm 0.04$) with a semi-major axis of a$_3 = 4.58 \pm 0.08$ au. Although such apparent changes in the orbital period could, in principle, result from apsidal motion, this is not expected for the circular orbit of DD~CrB. Magnetic activity is strongly disfavoured as the origin of the observed ETVs. The O–C deviations are well reproduced by an eccentric Keplerian model, while both the classical Applegate mechanism and the spin–orbit coupling model fail to meet the energetic requirements imposed by the observed modulation. 
Relatively large $\chi^2_\nu$ values around 3 are due to the scatter in the minima timings. Both our measurements and most of the mid-eclipse timings from the literature are based on the Kwee-van Woerden method \citep{kwee1956}, which is known to underestimate the measurement uncertainties. Although there had been no mention to the pulsations of the sdB component of DD~CrB, low-amplitude pulsations can affect minima profiles, and hence the mid-eclipse timings too \citep{2009A&A...505..239V}. While many minima timings from the literature have uncertainties of less than a second, their scatter is larger than a few seconds. Continued monitoring of DD~CrB is strongly encouraged to track the future evolution of the orbital period, which is suggested to be in an increasing trend by all of our models except for the combination of the linear, quadratic, and LiTE model.

\section*{Acknowledgements}

We gratefully acknowledge the support by The Scientific and Technological Research Council of Turkey (T\"UB\.{I}TAK) with the project 122F358. In this study, some of the photometric data obtained within the scope of project numbered 25ATUG100-3010 carried out using the TUG100 telescope at the TUG (T\"UB{\.I}TAK National Observatory, Antalya) site of the T\"urkiye National Observatories, and we would like to thank the T\"urkiye National Observatories and their personal for their valuable contributions. Some of the photometric data in this study were obtained with the T35 and T80 telescopes at the Ankara University Astronomy and Space Sciences Research and Application Center (Kreiken Observatory) with the project numbers 24A.T35.01 and 23A.T80.05, respectively. Some of the data presented in this paper were obtained from the Multimission Archive at the Space Telescope Science Institute (MAST). STScI is operated by the Association of Universities for Research in Astronomy, Inc., under NASA contract NAS5-26555. Support for MAST for non-HST data is provided by the NASA Office of Space Science via grant NAG5-7584 and by other grants and contracts. This research has made use of the NASA Exoplanet Archive, which is operated by the California Institute of Technology, under contract with the National Aeronautics and Space Administration under the Exoplanet Exploration Program. This work presents results from the European Space Agency (ESA) space mission Gaia. Gaia data are being processed by the Gaia Data Processing and Analysis Consortium (DPAC). Funding for the DPAC is provided by national institutions, in particular the institutions participating in the Gaia MultiLateral Agreement (MLA). The numerical calculations reported in this paper were partially performed at T\"UB{\.I}TAK ULAKB{\.I}M, High Performance and Grid Computing Center (TRUBA resources). UM acknowledges support from INAF 2023 MiniGrant funding program for the observations collected with Asiago Galileo 1.22m+B\&C and Copernico 1.82m+Echelle telescopes. The research of MW and PZ was partially supported by the project {\sc Cooperatio - Physics} of Charles University in Prague. The authors thank Jan Kára and Kamil Hornoch for their kind assistance with photometric observations and equipment at \ond\ observatory.
ACK gratefully acknowledges the financial support of the TÜBİTAK 2210 National Graduate Scholarship no. 1649B022401264, UK Science and Technology Facilities Council (STFC) and the Faculty of Natural Sciences, Keele University in the form of a PhD studentship.

\section*{Data Availability}
All light curves appearing for the first time in this article are presented as online material. Mid-eclipse timings derived from our own light curves as well as those of other observers' and TESS light curves are presented as online data sets too through Vizier Online.



\bibliographystyle{mnras}
\bibliography{dd_crb_final} 




\appendix

\section{Corner Plots}
\label{app:cornerplots}

\begin{figure*}
\includegraphics[width=0.95\textwidth]{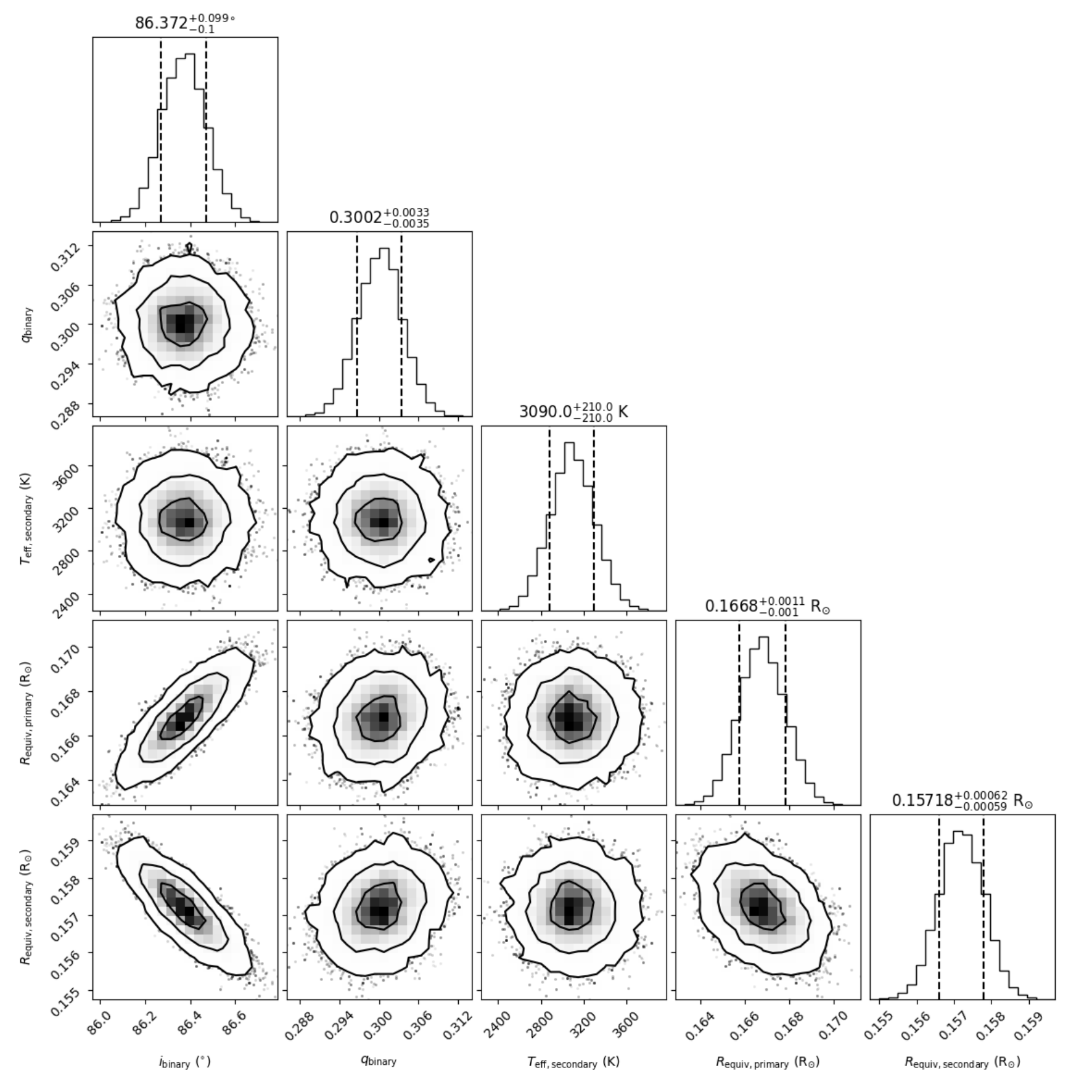}
\caption{Corner plot showing the posterior distributions of the model parameters in SDSS-u$^{\prime}$ filter and the correlations between these parameters.}
\label{fig:corner_sdssu}
\end{figure*}

\begin{figure*}
\includegraphics[width=0.95\textwidth]{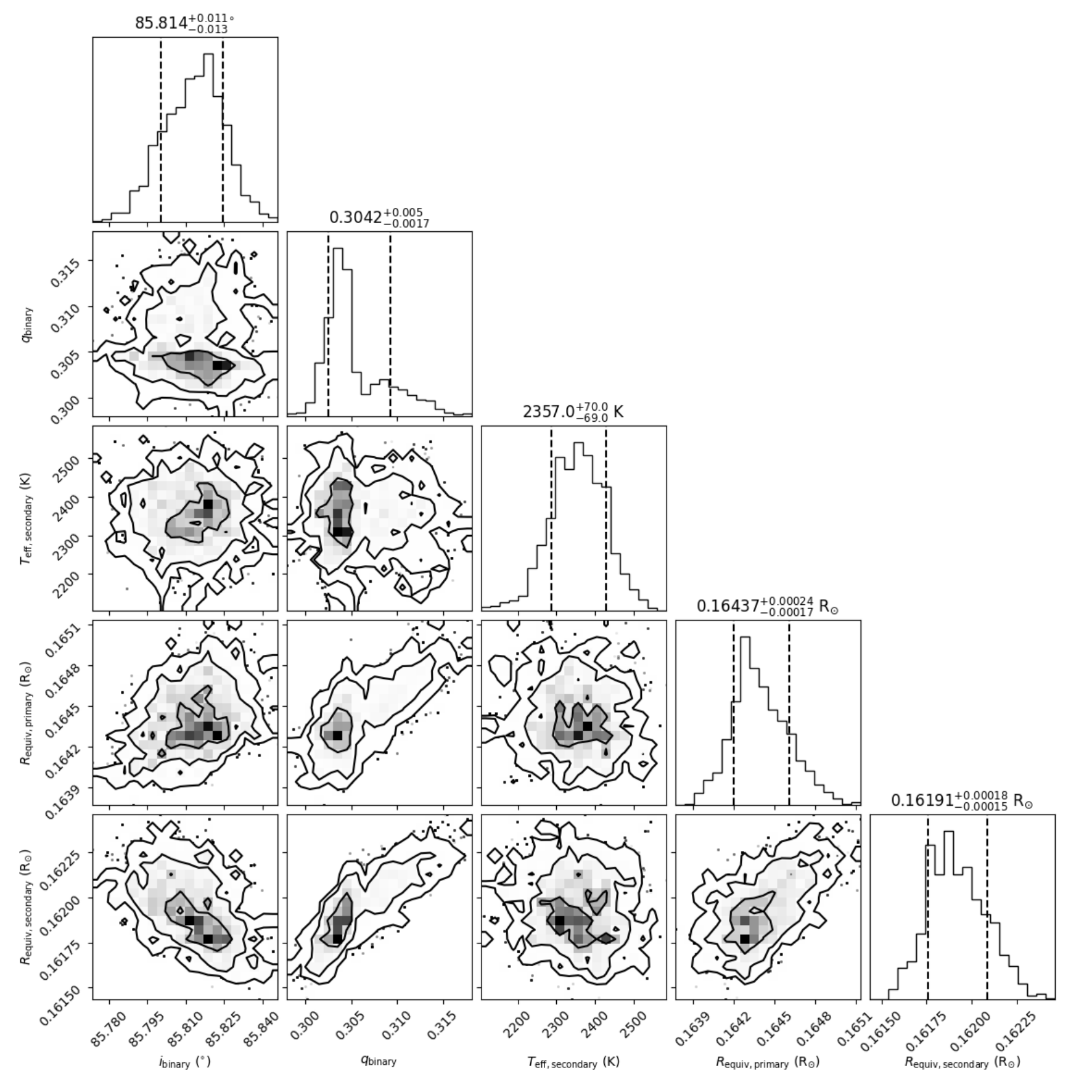}
\caption{Corner plot showing the posterior distributions of the model parameters from the simultaneous solution of all light curves achieved by fixing the albedo of the secondary to 1.0 and the correlations between these parameters.}
\label{fig:corner_allbands_A2fixed}
\end{figure*}

\begin{figure*}
\includegraphics[width=0.95\textwidth]{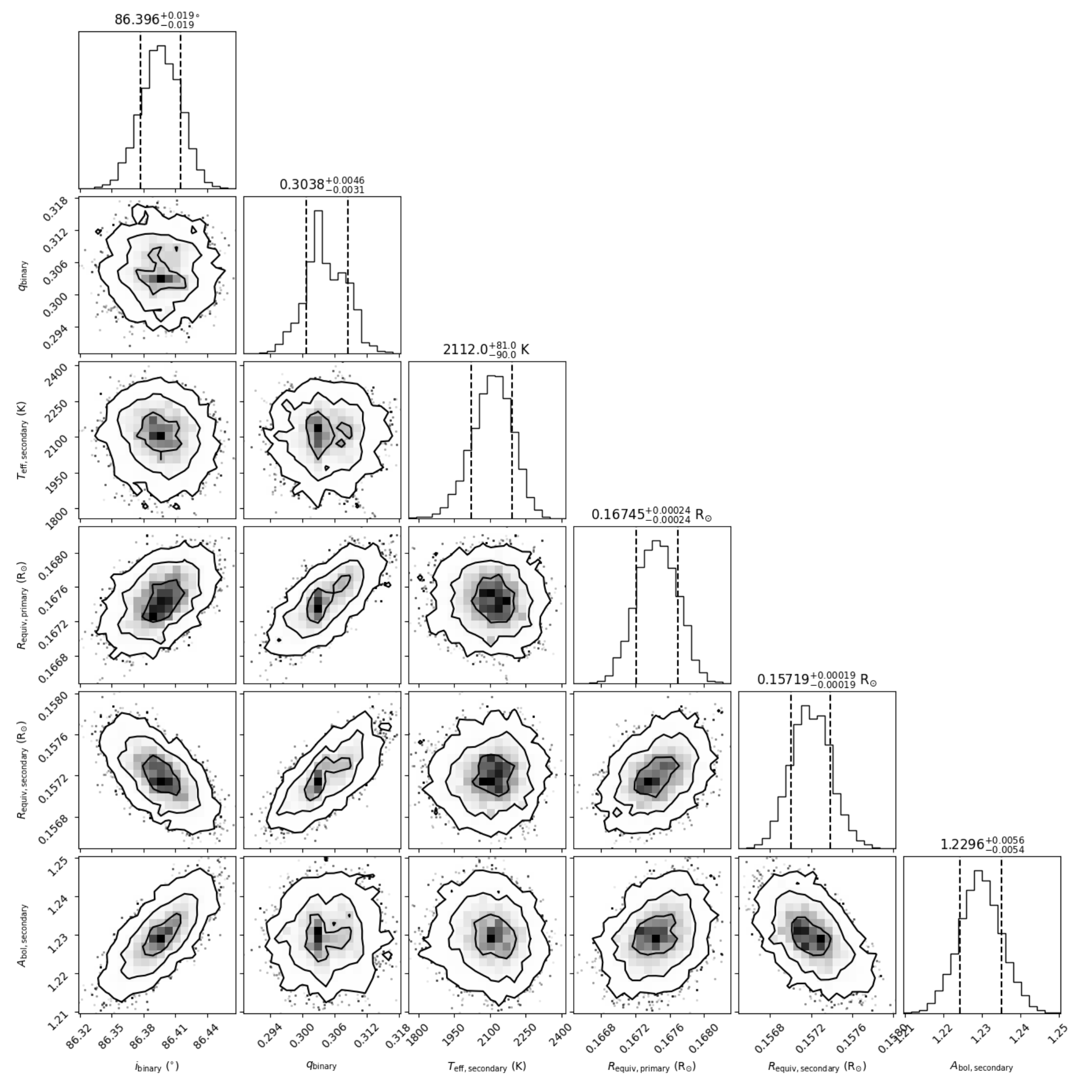}
\caption{Corner plot showing the posterior distributions of the model parameters from the simultaneous solution of all light curves achieved by adjusting the albedo of the secondary and the correlations between these parameters.}
\label{fig:corner_allbands}
\end{figure*}

\begin{figure*}
\includegraphics[width=0.95\textwidth]{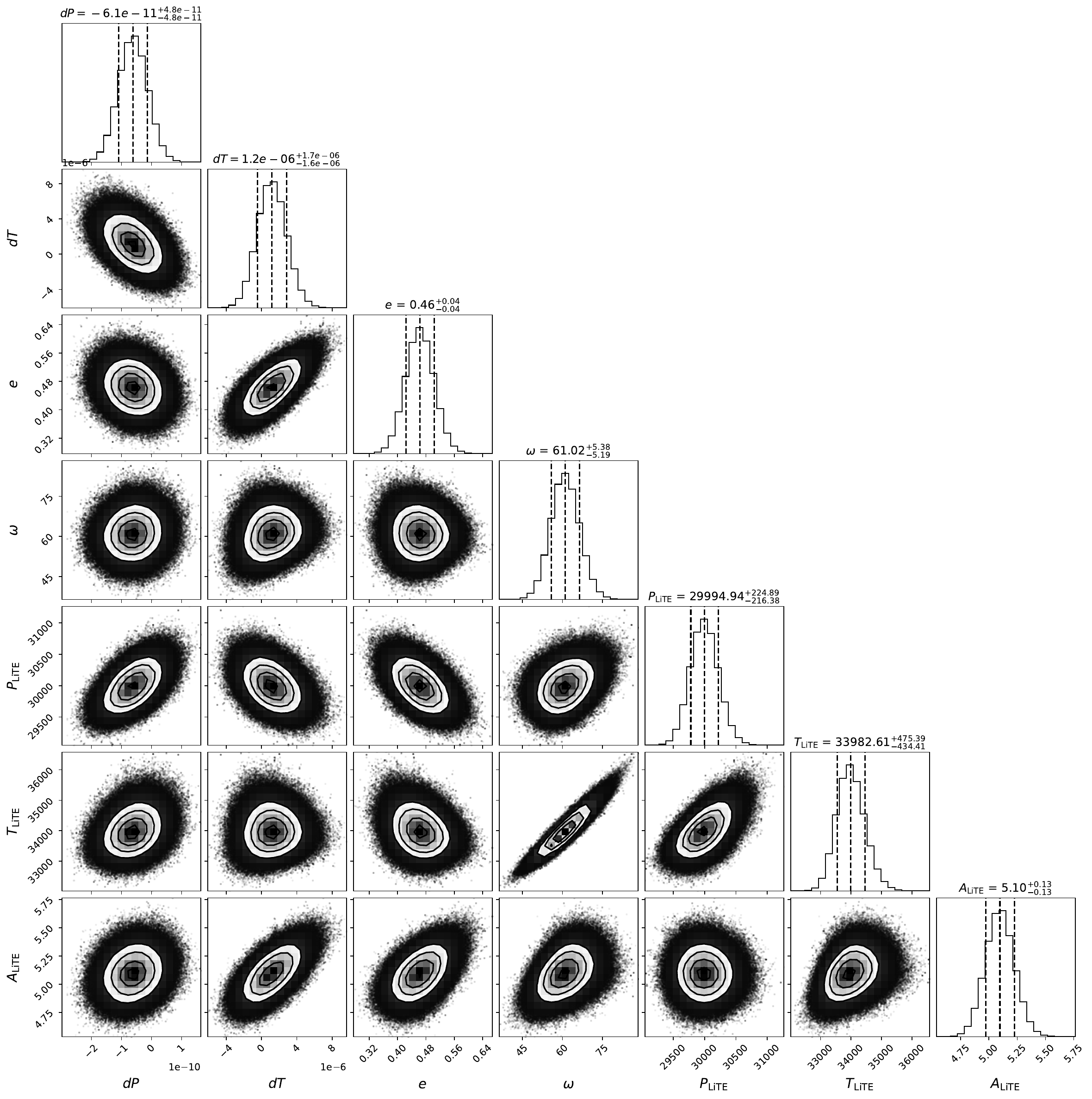}
\caption{Corner plot showing the posterior distributions of the Linear+LiTE model parameters and the correlations between these parameters.}
\label{fig:corner_etv}
\end{figure*}



\bsp	
\label{lastpage}
\end{document}